\documentclass{article}

\usepackage{arxiv}

\usepackage[utf8]{inputenc} 
\usepackage[T1]{fontenc}    
\usepackage{amsmath,amsthm}
\usepackage{amssymb}
\usepackage{hyperref}       
\usepackage{url}            
\usepackage{booktabs}       
\usepackage{amsfonts}       
\usepackage{nicefrac}       
\usepackage{microtype}      
\usepackage{cleveref}       
\usepackage{lipsum}         
\usepackage{graphicx}
\usepackage{natbib}
\usepackage{doi}

\usepackage{comment}
\usepackage{bm}
\usepackage{multirow}
\usepackage{setspace}
\usepackage{xcolor}
\usepackage{caption}
\usepackage{subcaption}
\usepackage{algorithm}
\usepackage{algpseudocode}

\newtheorem{definition}{Definition}
\newtheorem{lemma}{Lemma}

\newtheorem{theorem}{Theorem}
\newtheorem{proposition}{Proposition}
\newtheorem{example}{Example}

\title{Very large-scale neighborhood search for drone routing with energy replenishment}

\author{ \href{https://orcid.org/
0000-0002-0973-1817}{\includegraphics[scale=0.06]{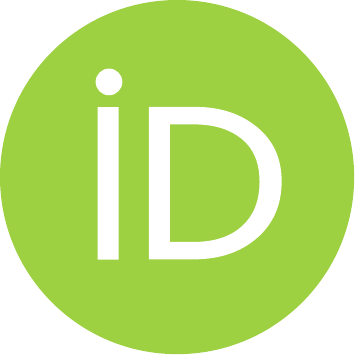}\hspace{1mm}Catherine ~Lorenz}\\
	Chair of Management Science / Operations \\ and Supply Chain Management, \\
	University of Passau,\\
	Dr.-Hans-Kapfinger-Stra{\ss}e 12,
	94032 Passau,\\
	\texttt{catherine.lorenz@uni-passau.de} \\
		\And
	\href{https://orcid.org/0000-0002-8926-3733}{\includegraphics[scale=0.06]{orcid.pdf}\hspace{1mm}Nicola~Mimmo} \\
	Department of Electrical, Electronic and \\
	Information Engineering and CIRI-ICT, \\
	University of Bologna, \\
Viale del Risorgimento 2, 
40136  Bologna\\
	\texttt{nicola.mimmo2@unibo.it} \\
	\And
	\href{https://orcid.org/0000-0002-6069-9330}{\includegraphics[scale=0.06]{orcid.pdf}\hspace{1mm}Alena~Otto} \\
	Chair of Management Science / Operations \\ and Supply Chain Management, \\
	University of Passau,\\
	Dr.-Hans-Kapfinger-Stra{\ss}e 12,
	94032 Passau,\\
	\texttt{alena.otto@uni-passau.de} \\
	\And
	\href{https://orcid.org/0000-0002-1499-8452}{\includegraphics[scale=0.06]{orcid.pdf}\hspace{1mm}Daniele~Vigo} \\
	Department of Electrical, Electronic and \\
	Information Engineering and CIRI-ICT, \\
	University of Bologna, \\
Viale del Risorgimento 2, 
40136  Bologna\\
	\texttt{daniele.vigo2@unibo.it} \\
}

\renewcommand{\shorttitle}{Drone routing with replenishment: Solution approaches}

\hypersetup{
pdftitle={Drone routing with replenishment: Solution approaches},
pdfsubject={Optimization},
pdfauthor={Catherine~Lorenz, Nicola~Mimmo, Alena~Otto, Daniele~Vigo},
pdfkeywords={very large-scale neighborhood search, drone routing with replenishment, dynamic programming},
}

\begin{document}
\maketitle

\begin{abstract}
	The Drone Routing Problem with Energy replenishment (DRP-E) belongs to a general class of routing problems with intermediate stops and synchronization constraints. In DRP-E, the drone has to visit a set of nodes and routinely requires battery swaps from a (potentially) mobile replenishment station. Contrary to widespread restrictions in the drone routing literature, several destinations may be visited in between two consecutive battery swaps. 
In this paper, we propose a nontrivial very large-scale neighbourhood for DRP-E, which synergetically leverages two large-sized polynomially solvable DRP-E \underline{s}ub\underline{p}roblems (SP1
and SP2). The number of feasible solutions in the resulting neighborhood is a multiple of those in
SP1 and SP2, and, thus, exponential in the input size of the problem,  
 whereas the computational time to search it remains polynomial. The proposed polynomial two-stage dynamic programming algorithm VLSN to search this neighborhood can be flexibly adjusted to the desired trade-off between accuracy and computational time. For instance, the search procedure can be converted into an exact algorithm of competitive runtime for DRP-E. 
In computational tests, the developed solution methods outperform current state-of-the art heuristics for DRP-E by a significant margin.  A case study based on a search for missing persons demonstrates that VLSN easily accommodates additional practice relevant features and outperforms the state-of-the-art solution in disaster relief by 20\%.
\end{abstract}

\keywords{very large-scale neighborhood search \and drone routing with replenishment \and dynamic programming}

\section{Introduction} \label{sec:intro}

Aerial drones are a highly promising modern technological invention that we experience up-close. Drones have infiltrated children's playrooms, fleets of logistics companies and ambitious speeches of operations and technology visionaries
. Just in the U.S. alone, around 850,000 drones are registered for commercial and recreational purposes \citep{faa2022}.

One of the main obstacles in the dissemination of drones, is the short battery life of commercially attractive small multi-rotor drones. Total airborne flight time  amounts to about 15 to 45 minutes depending on the size and the load of the drone \citep{liuetalunpublished}. Although a certain extension of endurance may come in the near future, e.g., with the forthcoming lithium-sulfur battery technology \citep{businesskorea2020}, many experts agree that a large number of commercially attractive applications will require periodic landings of the drone for battery recharges or swaps \citep{dronelife2017, 
tillemannandmccormick2018}. Current available technologies for ultra fast charging allow for the replenishment  of the drone's battery within about five minutes \citep{dronelife2020}. Even more impressive are so-called `hot' battery swapping techniques. The battery can be replaced automatically within just a few seconds, while the drone simultaneously remains fully powered \citep{ Liuetal2017}.

In this article, we study the scheduling of operations of a \textit{single drone}, in which the drone may have to replenish its energy regularly either by battery swapping or by recharging the battery to full capacity.  We formulate a basic optimization problem --\textit{the \underline{D}rone \underline{R}outing \underline{P}roblem with \underline{E}nergy replenishment (DRP-E)} (cf. Figure \ref{fig:instance}), which is relevant for a large number of applications.  The drone has to visit a set of \textit{points of interest (destinations)}. It can replenish its energy at one of the \textit{replenishment locations (RLs)} by visiting a replenishment station. These may be \textit{several stationary} replenishment stations. 
Alternatively, these may be possible rendezvous locations for the drone and one \textit{mobile station}. At such rendezvous locations, the vehicles can wait for each other and perform the replenishment operations safely. Specially-equipped trucks, replenishment robots, 
or autonomous platforms  
are examples of such mobile replenishment stations, which we call for short \textit{rover} in the following. The objective is to sequence the visits of destinations by the drone, schedule the drone's energy replenishment stops, and determine the locations of energy replenishment to minimize the \textit{makespan} subject to the following constraints:

\begin{figure}[t]
    \centering
    \includegraphics[scale= 0.33]{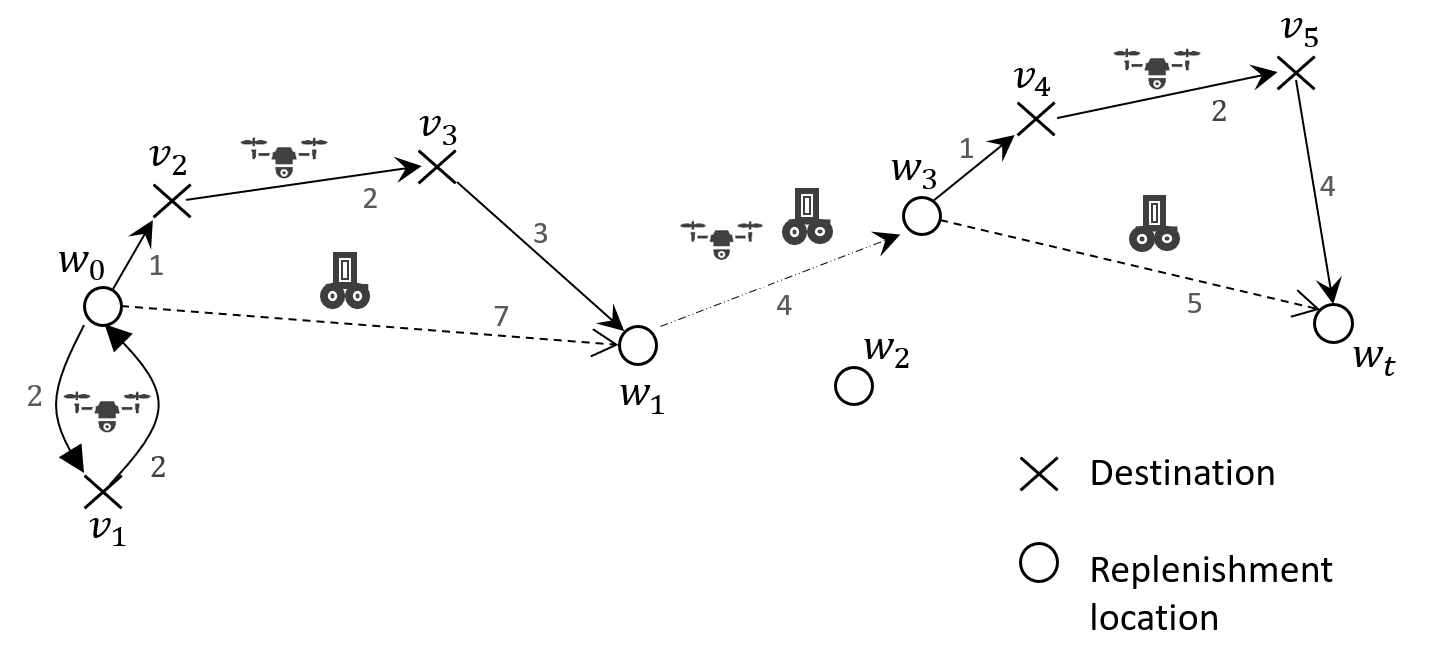}
    \caption{Illustration of a feasible solution for DRP-E: Route of the rover and the drone\\ \footnotesize{Example for $n_d=5$ destinations and $n_r=5$ replenishment locations (RLs), the makespan of the route equals 22. The route contains three operations with makespans $\mathcal{M}(o_1)=\mathcal{M}(w_0, v_1, w_0)=4$, $\mathcal{M}(o_2)=\mathcal{M}(w_0, v_2,v_3, w_1)=7$, and $\mathcal{M}(o_3)=\mathcal{M}(w_3, v_4, v_5, w_t)=7$ as well as one nonempty recharging leg with makespan $\mathcal{M}(w_1,w_3)=4$. The drone tour can be notated as an alternating sequence of operations and recharging legs $\pi_d=(r_0,o_1,r_1,o_2,r_2,o_3,r_3)$ by setting  $r_2=(w_1,w_3)$ and the recharging legs $r_0, r_1, r_3$ being empty.}} \label{fig:instance} 

\end{figure}


\begin{itemize}
\item The energy consumption of the drone between two subsequently visited RLs should not exceed the maximal energy capacity of the drone's battery;
\item In case of a rover, the replenishment starts when both the drone and the rover have reached the selected RL, in other words, \textit{waiting times} eventually emerge. 
\end{itemize}

The described problem setting, in which a single drone has to visit a set of nodes, emerges in a large number of applications \citep[see][]{ottoetal2018}. For example in disaster management, environmental monitoring, and infrastructure maintenance, the drone has to collect data from sensors \citep{sujitetal2013,jawharetal2014, turneretal2016,lietal2016b, albertetal2017}.  Drones also scan or take photographs of points of interest for maintenance \citep{guerreroandbestaoui2013}, police surveillance \citep{eversetal2014, zhangetal2015}, filming \citep{guerrieroetal2014} or for the purposes of precision agriculture \citep{tokekaretal2016}.  
Furthermore, the problem setting emerges in the surveillance of non-convex and/or not connected areas, where area discretization remains a state-of-the-art solution technique \citep{galceranandcarreras2013, cabreiraetal2019}.

DRP-E requires a particularly high degree of synchronization compared to many other routing problems with intermediate stops and synchronization constraints (cf. Section~\ref{sec:literature}). For such problems, solutions of construction heuristics of the route-first-split-second-type are notoriously hard to improve, notwithstanding the significant gap to optimality of these solutions in a number of realistic settings. Therefore, the focus of this paper on powerful and flexible improvement procedures for DRP-E appears to be a logical step towards achieving progress within this class of problems. 

In this article, we propose  efficient \textit{very large-scale neighborhood search (VLSN)} for DRP-E. VLSN is a two-stage approximate dynamic programming procedure{, in which the degree of approximation is controlled by a parameter $p$ (see Section~\ref{sec:vlns})}. VLSN possesses several attractive properties:
\vspace{-0.6cm}
\begin{quotation}
\item[\bf Efficiency:] VLSN examines an \textit{exponential} number of feasible solutions in  computational time which is polynomial in the number of destinations and RLs (but exponential in $p$).
\item[\bf Flexibility:] The size of the neighborhood and the computational complexity of the algorithm can be easily controlled by setting the value of the parameter $p$. VLSN can be also turned into an exact solution method with a competitive computational complexity (see Section~\ref{sec:exact}).
\item[\bf Generality:] VLSN uses a two-stage architecture with \textit{i)} a \textit{meta state graph} that subsumes the properties of the replenishment station and \textit{ii)} an \textit{operations state graph} that describes the properties of the drone. This architecture allows for easy accommodation of different drone and station characteristics without the need to adjust the overall logic of VLSN, such as limited communication range of the drone or limited energy capacity of the rover.  

\end{quotation}
To sum up, the outlined properties of VLSN makes it attractive for practitioners in a wide range of applications. In the following, we review related literature in Section~\ref{sec:literature} and conclude by stating the contribution of this article and providing the outline of the paper in Section~\ref{sec:contribution}.

\subsection{Literature review} \label{sec:literature}

The formulated DRP-E belongs to the rich class of node routing problems  \citep[see][for recent reviews]{Toth.2014,Braekers.2016,Konstantakopoulos.2020}. A characteristic feature of DRP-E is that the routes of the two involved vehicles -- the rover and the drone -- have to be \textit{synchronized} in space and time. In this regard, DRP-E resembles routing problems in two-echelon networks \citep{Cuda.2015,Guastaroba.2016,schifferetal2019,Accorsi.2020,Li.2021} --  
the Two-Echelon Vehicle Routing (2E-VRP) and Location Routing (2E-LRP) problems
as well as the Truck and Trailer Routing Problem (TTRP).
In DRP-E, the rover and the drone can be interpreted as first- and second-echelon vehicles, respectively. However, the routes of the heterogeneous vehicles in DRP-E are much more closely interrelated, and the synchronization is more complex:
For example, the drone can be launched in one RL, and rejoin the rover at another one, which may lead to mutual waiting times.  
Therefore, whereas state-of-the-art approaches for 2E-VRP, 2E-LRP and TTPR often rely on the separability of the routing sub-problems of the vehicles, this approach doesn't seem reasonable in our case and the solution method developed in this paper constructs the heterogeneous vehicle routes simultaneously.

A number of well-known routing problems are \textit{DRP-E special cases}. For example, the general class of asymmetric traveling salesman problems with replenishment arcs \citep{boland2000, makandboland2007}
can be converted to DRP-E using basic arc-node transformations and our solution method can be straightforwardly adapted. 
The traveling salesperson problem with hotel selection with the objective to minimize the number of nights spent in the hotel \citep{vansteenwegenetal2011, castroetal2014} is a DRP-E special case with stationary replenishment stations.

The literature on combined vehicle-drone operations is rapidly growing  \citep[see reviews of][]{ottoetal2018, Macrina.2020,Chung.2020}. Nevertheless, most of such articles describe specific delivery applications, in which the drone delivers packages alongside the delivery truck, but cannot carry more than one package at a time, and must rejoin the truck after every single delivery  \citep[cf.][]{murrayandchu2015,bouman2017,Freitas.2019,Roberti.2021}. A few articles on drone routing consider special cases of DRP-E by introducing additional constraints or relaxing the temporal synchronization of the vehicle routes \citep{luoetal2017,Luo.2018, mathewetal2015, Wang.2020}. Some other articles on combined vehicle-drone operations share certain properties with DRP-E, but in fact, can be considered a separate problem due to their application-specific components \citep{karakandabdelghany2019, yu2018}. 

To the best of our knowledge, only \citet{poikonenandgolden2020} consider the problem setting described in DRP-E. The authors developed a route-first-split-second construction heuristic, dubbed \textit{Route, Transform, Shortest Path (RTS)}, which optimally inserts replenishment stops within a \textit{fixed} sequence of destinations (= `splits' this sequence). \citet{poikonenandgolden2020} proceed by showing how to adapt the RTS heuristic to the cases of several homogeneous drones, and a realistic energy drain function. As is shown in Section~\ref{sec:performance}, our solution procedure VLSN dominates RTS, since all the solutions implicitly enumerated by RTS are also examined by VLSN.

\subsection{Contribution and the outline of the paper} \label{sec:contribution}

This paper is one of the first articles in the literature on DRP-E. As DRP-E involves a close temporal synchronization of two vehicles, it is challenging to solve, and the solutions of the state-of-the-art route-first-split-second heuristics are hard to improve, although the remaining gap to optimality may be significant  (cf. also Section~\ref{sec:casestudy}). Moreover, DRP-E is a general problem formulation that applies to many real-world applications and is a generalization of several well-known optimization problems. 

Our paper makes the following contributions:
\begin{itemize}
\item It proposes a very large-scale neighborhood search for DRP-E, which examines an exponential number of feasible solutions efficiently. 
The developed very large-scale neighborhood approach \citep[see][for an overview]{ahuja.2002} is nontrivial as it permits synergetic leverage of two large-sized polynomialy solvable DRP-E \underline{s}ub\underline{p}roblems (SP1 and SP2). Roughly speaking,
the number of examined feasible solutions is a multiple of those in SP1 and SP2, whereas the computational time remains polynomial (cf. discussions in Section~\ref{sec:vlns}).
\item The paper suggests an exact solution procedure with competitive computational complexity, which is the first exact procedure for DRP-E proposed thus far.  
\item It proposes a structured benchmark data set for DRP-E.
\item Computational experiments demonstrate that VLSN significantly outperforms the state-of the-art route-first-split-second heuristic of \cite{poikonenandgolden2020} for DRP-E in terms of solution quality. This
 is especially significant, since the results of this type of heuristic are hard-to-improve otherwise, e.g. by embedding it into common metaheuristic frameworks.
\item In a real-world case study on the search of missing persons we show, that VLSN can be easily adapted to incorporate rich application specifics. It achieves a 54-minute reduction in the makespan of a 4.5-hour search mission received by the state-of-the-art solution in disaster relief. 
\end{itemize}

In the following, we formulate DRP-E in Section~\ref{sec:problem}. Section~\ref{sec:vlns} outlines VLSN. Section~\ref{sec:exact} 
discusses how to turn VLSN into an exact solution procedure.  We proceed with extensive computational experiments in Section~\ref{sec:computational}. Section~\ref{sec:conclusion} concludes the paper with final remarks and an outlook. 

\section{The drone routing problem with energy replenishment} \label{sec:problem}

To avoid distracting details in outlining the mechanics of VLSN, we focus on a basic formulation for DRP-E, which we introduce formally in Section~\ref{sec:formal}. \textit{W.l.o.g.}, we restrict our discussion to the more general case of a mobile replenishment station (rover).  In Section~\ref{sec:practical}, we discuss the validity of stated assumptions under the context of realistic applications.

\subsection{Problem formulation} \label{sec:formal}

%
DRP-E can be described as follows.
\begin{description}
\item[Given:] 
\end{description}
\begin{itemize}
\setlength{\itemsep}{0cm}
\item set of destinations $V_d$, $|V_d|=n_d$, 
\item set of RLs $V_r$, $|V_r|=n_r$, 
\item special RLs $w_0\in V_r$ (initial depot) and $w_t\in V_r$ (target depot),
\item 
drone flight times  $c_d: (V_d\cup V_r) \times (V_d \cup V_r) \rightarrow \mathbb{R}_{\geq 0}$, 
\item 
rover travel times  $c_r: V_r \times V_r \rightarrow \mathbb{R}_{\geq 0}$,
\item maximal flight time $e_{max}\in\mathbb{R}_{> 0}$ enough to visit each destination in a return flight from its nearest RL;  $e_{max}$ is computed from the energy capacity of the drone's battery.
\end{itemize} 


We define $c_d$ and $c_r$ as the \textit{shortest} drone flight and rover travel times between a pair of locations, respectively, thus the triangular inequalities are valid, i.e. $c_d(i,j)\leq c_d(i,k)+c_d(k,j) \ \forall k \in V_d\cup V_r$ (equivalently for $c_r$).

We call \textit{operation} a drone sortie starting from a RL, then visiting at least one destination and returning to the same or a different RL (see Figure \ref{fig:instance}). We notate operation $o$ with $k(o)$ visited destinations as an ordered sequence $o=w,v_{O1},v_{O2},...v_{Ok},w'=wsw'$, with $w, w' \in V_r$ being the starting and ending RLs of the operation, respectively, 
$Oi$ denoting the index of the $i$th destination in the operation,
and $s$ being the destinations sorted according to their visiting order.
We call operation $o$ \textit{feasible}, if the drone flight time $C_d$ does not exceed $e_{max}$, i.e.,
\begin{center}
$C_d(o):=c_d(w,v_{O1})+\sum_{i=1}^{k(o)-1} c_d(v_{Oi},v_{O(i+1)})+c_d(v_{Ok},w')\leq e_{max}$.
\end{center}

We call \textit{recharging leg} $r$ a sequence of $k(r)$ consecutive RLs, where the drone travels on the back of the rover without spending energy,  $r=(w_{R1},\ldots, w_{Rk})$ with $w_{R1},\ldots, w_{Rk}\in V_r$.

\begin{description}
\item  \textbf{Objective:} Determine a feasible \textit{drone tour}, i.e. a sequence of  operations and recharging legs, $\pi_d=(r_0, o_1, r_1, o_2,\ldots)$ such that
\begin{itemize}
\item Each destination is visited, i.e. $v\in \pi_d$ $\forall v\in V_d$.
\item Sequence $\pi_d$ starts at the initial depot $w_0$ and ends at the target depot $w_t$.
\item Each operation $o\in \pi_d$ is feasible. 
\item The makespan $\mathcal{M}(\pi_d)=\sum_{o \in \pi_d}\mathcal{M}(o)+\sum_{r \in \pi_d}\mathcal{M}(r)$ is minimized, where
\begin{itemize}
\item $\mathcal{M}(o)=\max\{C_d(o);c_r(w,w')\}$, for $o=wsw'$ (makespan of operation $o$).
\item $\mathcal{M}(r)=\sum\limits_{i=1}^{k(r)-1} c_r(w_{Ri},w_{R(i+1)})$ for $r=(w_{R1},\ldots, w_{Rk})$ (makespan of recharging leg $r$).
\end{itemize}
\end{itemize}
\end{description}
Observe that the makespan of a recharging leg solely depends on the rover, because the drone travels on its back without spending any energy. On the other hand, the makespan of an operation equals to the maximum of the drone's and the rover's travel times, because both vehicles have to meet for the battery exchange. 
See Figure~\ref{fig:instance} for an example.
W.l.o.g., we use the following properties of DRP-E in our solution procedure VLSN, which directly follow from the triangular inequalities of the drone flight times and rover travel times
:
\begin{itemize}
\item[1)] 2-node recharging legs:  $k(r)=2$ $\forall r\in \pi_d$ (trivial recharging legs $r=(w,w)$ for $w\in V^r$ are possible)
\item[2)] Each destination is visited exactly once.  
\end{itemize}

DRP-E is NP-hard, as its special case with unlimited battery capacities of the drone reduces to a standard traveling salesman problem. 

\subsection{Discussion of assumptions} \label{sec:practical}

Let examine the problem assumptions and how the proposed approach VLSN can be extended to more complicated problem settings, in which some formulated assumptions are relaxed.

\begin{quotation}
\vspace{-0.5cm}
\item [\bf Mobile replenishment station.] We can apply our method to the case of stationary replenishment stations by setting the travel times of the station to zero and prohibiting the so-called recharging legs, during which the drone uses the station as a transportation mule.


\item[\bf Energy consumption proportional to the flight time.] We assume that the energy expenditure rate of the drone is constant and that it is proportional to its (constant) velocity, which applies if the drone flies at about the same altitude (cf. Section \ref{sec:casestudy}). To accommodate more complicated energy expenditure functions, the value function in the operations state graph turns out to be multi-valued, since we have to track both the elapsed time, and the consumed energy. By using labeling algorithms similar to \citep{smith&boland2012}, our VLSN algorithm can be adjusted straightforwardly to handle this feature.
  
\item[\bf Energy replenishment to full battery capacity.] This assumption corresponds with the replenishment performed via battery swaps. The relaxation of this assumption can be addressed by extending the definition of states in the meta state graph by including a discretization of the drone's energy level, \citep[see e.g.][]{yu2018}. 

\item[\bf Deterministic parameters.] We assume that all parameters are deterministic. Stochastic factors, such as wind gusts, may play an important role in some applications. We leave this element for further research.

\end{quotation}

\section{The very large-scale neighborhood search} \label{sec:vlns}
We propose a very large-scale neighborhood that synergetically and nontrivially unites two polynomially solvable DRP-E \textit{subproblems} (SPs). 
The first one, SP1, is a traveling salesman problem over destinations with restrictions on the relative positions of the nodes as explored in the seminal studies of \citet{balas1999} and \citet{balasandsimonetti2001}.
The second one, SP2, is DRP-E with a fixed sequence of destination visits $x$. As shown by \citet{poikonenandgolden2020}, for the given $x$, we can schedule replenishment stops and determine their locations optimally 
in $O \left(n^2_dn^2_r\right)$ time.

Each of these subproblems has exponentially many feasible solutions. Our neighborhood combines these polynomially solvable subproblems into one: For \textit{each} sequence of destinations in SP1, all possible insertions of RLs are examined. As a result, the number of feasible solutions is a \textit{multiplication} of those in SP1 and SP2, if we place some of the limited energy capacity subtleties aside (see Theorem~\ref{theo: size_BS-R}). Contrarily,  the computational complexity to search this neighborhood remains polynomial (see Theorems~\ref{theo: size_BS-R} and \ref{theo:complexity}). 

We call the designed neighborhood \textit{\underline{B}alas-\underline{S}imonetti neighborhood with \underline{R}eplenishment (BS-R)}, or $\mathcal{N_{BS-R}}(x,p)$ in mathematical notation. To search this neighborhood, we propose a two-stage dynamic programming procedure VLSN consisting of an \textit{operations state graph (ops graph)} and a \textit{meta state graph (meta graph)}. The meta graph explores how to construct an optimal drone tour in BS-R by combining operations and recharging legs. 
For each subset of destinations and a pair of start- and end-RLs, 
the ops graph computes a feasible operation for BS-R neighbors with a minimum required drone flight time. Much of mathematical effort is devoted to keeping these graphs polynomial in size.

Throughout this section, we assume that $x$ denotes a Hamiltonian path over $V_d$ and that destinations in $x$ are enumerated in the increasing order of their position, i.e. $x=(v_1, v_2, \ldots, v_{n_d})$. We also use shortcut $[n]$ for the set of integers $\{1,\ldots,n\}, n\in\mathbb{N}$, and similarly $[a_1,b_1]$ for integers $\{a_1,\ldots,b_1\}$. The position of destination $v_i$ in a permutation $\sigma$ of the visiting sequence $x$ is referred to as $\sigma(i)$, while $\sigma^{(-1)}(l)$ returns the destination at the $l$th position in permutation $\sigma$.

We proceed with the formulation of BS-R and its properties in Section~\ref{sec:balas}, which also summarizes the main results of this work. Then, Sections~\ref{sec:operations} and~\ref{sec:meta} explain the construction of the ops graph and the meta graph, respectively. We conclude with the discussion of a special case in Section~\ref{sec:outline}, in which the drone tour starts and ends in the same location $w_0=w_t$.

\subsection{Balas-Simonetti neighborhood with replenishment} \label{sec:balas}
Before defining the proposed BS-R, let discuss the classical \textit{Balas-Simonetti neighbourhood (BS)}. 

For a given Hamiltonian path of destinations $x:(v_1,...,v_{n_d}), v_i \in V_d$, and a parameter $p \in \mathbb{N}$, the Balas-Simonetti neighbourhood $\mathcal{N_{BS}}(x,p)$ is a collection of permutations defined by the following precedence constraints:
\begin{align}
&x'= (v'_1,v'_2,...v'_{n_d})=(v_{\sigma^{-1}(1)},v_{\sigma^{-1}(2)},...v_{\sigma^{-1}({n^d})}) \in \mathcal{N_{BS}}(x,p) \text{ iff} \nonumber\\
&  \forall i,j \in [{n_d}] \text{ s.t. } i + p \leq j \text{ we have } \sigma(j)  > \sigma (i) 
\label{eq:bs}
\end{align}

\begin{table}[htp]
	\caption{Balas-Simonetti neighbourhood for $x=(v_1,v_2, v_3, v_4,v_5)$ and $p=2$
	 } \label{tab:Table_BSexample}
	 \begin{center}
	\begin{tabular}{p{10.5cm}p{5cm}}
		\toprule
	\multirow{6}{*}{\begin{minipage}{.45\textwidth}
      \includegraphics[scale=0.70]{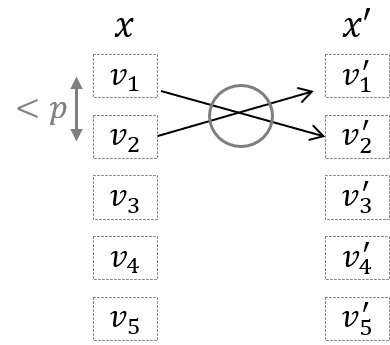}
    \end{minipage}}	
		& Neighbors in $\mathcal{N_{BS}}(x,2)$: \\ 
		&  $(v_1, v_2, v_3, v_4, v_5)$  \\ 
		& $(v_1, v_2, v_3, \bm{v_5, v_4})$ \\
		& $(v_1, v_2, \bm{v_4, v_3}, v_5)$  \\
		& $(v_1, \bm{v_3, v_2}, v_4, v_5)$   \\
		& $(v_1, \bm{v_3, v_2, v_5, v_4})$  \\
		& $(\bm{v_2, v_1}, v_3, v_4, v_5)$  \\
			& $(\bm{v_2, v_1}, v_3, \bm{v_5, v_4})$  \\ 
			& $(\bm{v_2, v_1, v_4, v_3}, v_5)$  \\
			& \\
			& \\
		\midrule[\heavyrulewidth]
	\end{tabular}\\
	\footnotesize{The table on the right lists all the neighbors in $\mathcal{N_{BS}}(x,p)$. The figure on the left illustrates, how to construct  these neighbors. So-called matching arcs depict, to which position a node in $x$ moves in $x'$
		. As shown in the left figure, to receive a valid $\mathcal{N_{BS}}(x,p)$-neighbor, only matching arcs that start in nodes $v_i,v_j: |i-j|<p$ may cross.}
\end{center}
\end{table}

Table~\ref{tab:Table_BSexample} illustrates how to construct BS-neighbors for $n^d=5$ and $p=2$. 

Now, we formally define BS-R, which integrates the replenishment decisions.

\begin{definition}[\textbf{Balas-Simonetti neighbourhood with replenishment $\mathcal{N_{BS-R}}(x,p)$}]\label{def:bsr}
For a given Hamiltonian path of destinations $x$ and a given parameter $p \in \mathbb{N}$, we define the \textit{Balas-Simonetti neighbourhood with replenishment} $\mathcal{N_{BS-R}}(x,p)$ as the subset of all 
drone tours $\pi_d$ such that the drone tour restricted to destinations $\pi_d(V_d)$ respects the precedence constraints of $\mathcal{N_{BS}}(x,p)$:
\begin{align}
\pi_d \in \mathcal{N_{BS-R}}(x,p) \text{ \ iff \ } \pi_d(V_d) \in \mathcal{N_{BS}}(x,p) \nonumber
\end{align}  
\end{definition}

Observe that we can initialize neighborhood BS-R without having a feasible DRP-E solution, since it is defined just on the visiting sequence of destinations $x$ and parameter $p \in \mathbb{N}$.

Having defined BS-R, we are able to state the main results of this work.

\begin{theorem}\label{theo: size_BS-R}
For a given sequence $x$ of destinations and a given parameter $p$, the number of feasible drone tours that belong to $\mathcal{N_{BS-R}}(x,p)$ for DRP-E  grows exponentially with the input size $n_d$ (number of destinations). 
\begin{flalign}
\text{\textit{If the maximal flight time is unlimited, }}e_{max}\rightarrow\infty\text{\textit{,~~~}}& \vert \mathcal{N_{BS-R}}(x,p) \vert \geq (n_r)^{2n_d}\cdot \left(\frac{p-1}{e}\right)^{n_d-1} &\label{eq:numseq1}\\
\text{\textit{Otherwise, for }}e_{max}\in \mathbb{R}\text{\textit{,~~~~~~~~}}&\vert\mathcal{N_{BS-R}}(x,p) \vert \geq \left(\frac{p-1}{e}\right)^{n_d-1} &\label{eq:numseq2}
\end{flalign}
\end{theorem}
\proof
Consider the case of $e_{max}\rightarrow\infty$ first. By Proposition 1 in \citet{balasandsimonetti2001}, $\mathcal{N_{BS}}(x,p)$ contains at least $\left(\frac{p-1}{e}\right)^{n_d-1}$ sequences of destinations. For each sequence of destinations $x'$, VLSN implicitly examines all possible replenishment schedules and inserts replenishment stops optimally. By Property 1) in Section~\ref{sec:formal}, it is enough to examine two-node recharging legs between each pair of operations. I.e., after each destination in $x'$, we can either fly to the next destination or end the current operation and insert a recharging leg, and there are $n_r^2$ possible alternative recharging legs, except for the last one that should end in $w_t$. We add $n_r$ variants for the initial recharging leg that starts in $w_0$. This results in $n_r\cdot (n_r^2+1)^{(n_d-1)}\cdot n_r$, or $O\left(n_r^{2n_d}\right)$, replenishment schemes for each examined sequence of destinations. Lower bound~(\ref{eq:numseq1}) follows immediately.

If $e_{max}$ is limited, then we can construct at least one feasible drone tour for each sequence of destinations $x'=(v_{\sigma^{-1}(1)},v_{\sigma^{-1}(2)},...v_{\sigma^{-1}(n_d)})\in \mathcal{N_{BS}}(x,p)$. Indeed, in Section~\ref{sec:formal} we assumed that the maximal flight time $e_{max}$ is large enough to  visit each destination in a return flight from its nearest RL. Therefore at least the tour, where the drone replenishes energy after each visited destination in the closest-by RL, is feasible for each sequence of destinations $x'$. 
Thus, the number of examined feasible drone tours is not less than the number of neighbors in $\mathcal{N_{BS}}(x,p)$. 
\endproof

We note that the bound in Theorem~\ref{theo: size_BS-R} is overly pessimistic, since the actual number of the examined drone tours is many times larger than this bound and rapidly grows with $n^d$ even at such low values of $p$ as $p=2$, as we show in computational experiments in Section~\ref{sec:performance}.

\begin{theorem}\label{theo:complexity}
For any given instance of DRP-E, sequence $x$ of destinations and given $p$, a best feasible drone tour $\pi_d^*\in \mathcal{N_{BS-R}}(x,p)$ can be found in the worst-case runtime of  $O(p^2 {n}^2_r {n}^2_d \cdot 4^{p})$.
\end{theorem}
\proof
The time complexity of VLSN  sums up from the  complexities to construct and compute the ops graph at the first stage and the meta graph at the second stage, which are $O\left(4^{p}\left(n_r^2n_d^2p+n_r n_d^2 p^2\right)\right)$ (see Section~\ref{sec:ops_complexity}) and $O(n^2_d n_r^2\cdot 4^{p})$ (see Section~\ref{sec:meta_complexity}), respectively.  This results in $O(p^2 {n}^2_r {n}^2_d \cdot 4^{p})$ for VLSN. 
\endproof

Computational experiments of  Section~\ref{sec:performance} show that the bound in Theorem~\ref{theo:complexity} is pessimistic in case of a realistic battery capacity of the drone: With each increase of $p$ by 1, the observed runtime of VLSN about doubles.

\subsection{Operations state graph} \label{sec:operations}

In the ops graph, we generate a collection of operations by computing the best way to fly over any subset of destinations $S \subseteq V_d$ starting and ending in any pair of RLs $w$ and $w'$, respectively. The challenge is to consider only those operations which are relevant for the BS-R neighbors of $x$ given neighborhood $\mathcal{N_{BS-R}}(x,p)$, i.e. only those which can be found in at least one neighboring drone tour $\pi_d$ of $\mathcal{N_{BS-R}}(x,p)$. We call such operations $\mathcal{N_{BS-R}}$\textit{-valid}.
\begin{figure}[!htb]
\centering
\includegraphics[scale=0.7]{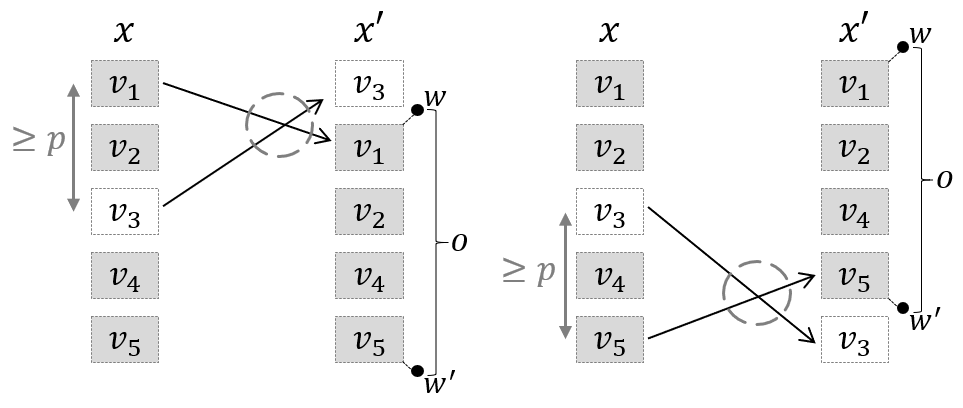}
\caption{Example of a $\mathcal{N_{BS-R}}$-invalid operation $o=wsw', s=(v_1,v_2,v_4,v_5)$ for $p=2$ \\ \footnotesize{Placing destination $v_3 \notin s$ before operation $o$ in a drone tour $\pi_d$ forces the crossing of the matching arrows from $v_3$ and $v_1$, with $\vert 3-1 \vert \geq p$. Placing destination $v_3 \notin s$ after operation $o$ in a drone tour $\pi^d$ forces the crossing of the matching arrows from $v_3$ and $v_5$, with $\vert 3-5 \vert \geq p$. This implies a violation of (\ref{eq:bs}). }}\label{fig:invalid_op}
\end{figure}

We have to recognize $\mathcal{N_{BS-R}}$-valid operations efficiently already during the construction of the ops graph, to ensure that the algorithm remains polynomial. However, it is a nontrivial task, since the sequence of destinations $s$ of some operation $wsw'$ may force \textit{the remaining operations} to violate  the Balas-Simonetti precedence relations~(\ref{eq:bs}). For instance in Figure~\ref{fig:invalid_op}, sequence of destinations $s=(v_1, v_2, v_4, v_5)$ does not violate  (\ref{eq:bs}) for $p=2$ itself. Nevertheless, operation $o=wsw'$ is not $\mathcal{N_{BS-R}}$-valid, i.e. no BS-R neighbor may contain this operation. This is because node $v_3$, which is excluded from $o$, cannot be assigned to any other operation without violating (\ref{eq:bs}).

In Sections~\ref{sec:ops_general} and \ref{sec:ops_complexity}, we explain how to construct the ops graph in case of $e_{max}\rightarrow\infty$ and discuss its time complexity. Afterward, we show how to respect limited $e_{max}$ in Section~\ref{sec:feasible_ops}.




\subsubsection{Architecture of the ops graph for unlimited flight time} \label{sec:ops_general}

\begin{figure}
\centering
\includegraphics[scale=0.53]{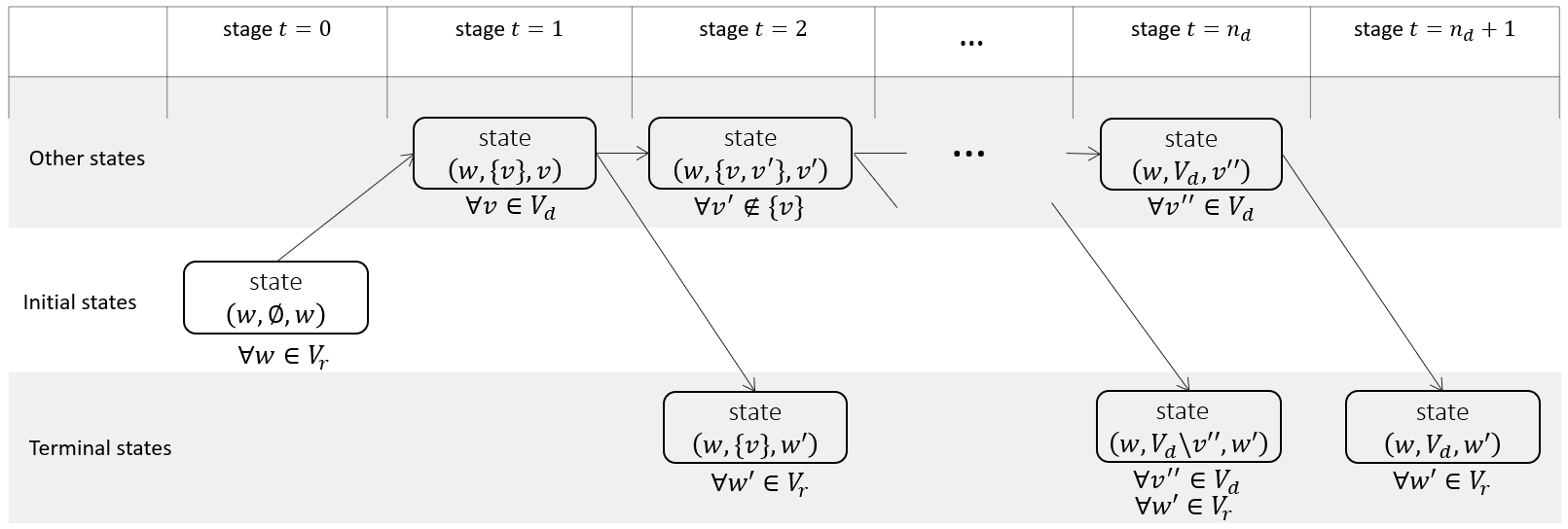}
\caption{A schematic illustration of the ops graph} \label{fig:operationsgraph}
\end{figure}

\textit{States} in ops graph $\mathcal{O}$ are arranged in stages $t\in\{0,1,...,n_d+1\}$ and have the form $(w,S,v)$ with $w\in V_r$ being the initial RL, $S\subseteq V_d$ being the set of destinations of the current operation visited so far and $v\in S\cup V_r$ referring to the current position of the drone (see Figure~\ref{fig:operationsgraph}). Stage 0 consists of $m$ initial states $(w, \emptyset, w), w\in V_r$ indicating the initial RL of the drone. State $(w,S,v)$ with $v\in S$ belongs to stage $|S|$. Terminal states $(w,S,w')$  with $w,w'\in V_r$ and $S\subseteq V_d, S \neq \emptyset$ correspond to completed operations and belong to stage $|S|+1$. 

By analogy with $\mathcal{N_{BS-R}}$-valid operations, we define $\mathcal{N_{BS-R}}$-valid states, which are 
\begin{itemize}
\item all initial states, 
\item terminal states $\left(w,S,w'\right)$ such that there exists at least one associated $\mathcal{N_{BS-R}}$-valid operation $o=wsw'$ with $\{s\}=S$, 
\item and all other states $\left(w,S,v\right)$, for which there exists at least one associated $\mathcal{N_{BS-R}}$-valid operation $o=wsw'$, $\{s\}=S$, with a visiting sequence of destinations $s$ terminating in $v\in S$. 
\end{itemize}
We construct the ops graph such that $\mathcal{O}$ contains \textit{only} $\mathcal{N_{BS-R}}$-valid states.

Transition arcs in $\mathcal{O}$ connect states between consecutive stages. Outgoing arcs from initial states $(w, \emptyset, w)$ income in states $(w, \{v\}, v)$ with $v\in V_d$ and have the cost of $c_d(w,v)$. Non-terminal states $(w,S,v)$ with $v\in S$ are adjacent to the following groups of states:
\begin{itemize} 
\item $(w,S\cup\{v'\},v')$  with $v'\in V_d\setminus S$, and $v'\notin S$; the cost of the corresponding arcs equals $c_d(v,v')$ (the prolongation of the drone flight by one not yet visited destination),
\item $(w,S,w')$ with $w'\in V_r$; the cost of the respective arcs equals $c_d(v,w')$ (closure of the operation).
\end{itemize}

Figure \ref{fig:operationsgraph} depicts the overall architecture of the described state graph. 

In Lemma~\ref{lem:valid_trans}, we explain, how to construct outgoing transition arcs from some $\mathcal{N_{BS-R}}$-valid state efficiently, such that the head state of the transition arc is an $\mathcal{N_{BS-R}}$-valid state as well. Since all the initial states $(w, \emptyset, w), w\in V^r,$ as well as states on the first stage are $\mathcal{N_{BS-R}}$-valid by definition, this lemma essentially explains, how to construct the ops graph efficiently in the forward direction. 

\begin{lemma} \label{lem:valid_trans}
For the BS-R neighborhood $\mathcal{N_{BS-R}}(x,p)$ with $p\in \mathbb{N}$, consider non-terminal $\mathcal{N_{BS-R}}$-valid state $(w,S,v)\in \mathcal{O}$ with $|S|\geq 1$. Consider numbers $m:=\min\{j:v_j\in S\}$  and $M:=\max\{j:v_j\in S\}$. Then,
a corresponding state $(w,S\cup\{v'\},v'=v_i)$ with $v'\in V_d\setminus S$ is $\mathcal{N_{BS-R}}$-valid iff
\begin{align}
\begin{cases} i\in[M-p+1, \max\{M-1, m+2p-1\}] & \text{ or }\\ i\in[\max\{M+1, m+2p\},M+p] \text{ and } \left(v_j\in S \text{ } \forall j\in[m+p,i-p]\right) \end{cases} \label{eq:destseq}
\end{align}
Each corresponding terminal state $(w,S,w')$ with $w'\in V_r$ is $\mathcal{N_{BS-R}}$-valid.

\end{lemma}

\proof
See Appendix~\ref{sec:constr_ops_graph}. 
\endproof

We compute all shortest paths between initial and terminal states in $\mathcal{O}$ by applying the following Bellman's equation in the forward induction manner for states $(w,S,v)\in\mathcal{O}, \ v\in S \cup V_r$:
\begin{align} 
\zeta(w,S,v)=\begin{cases}
 c_d(w,v) & \text{if } |S| \in \{0,1\}  \\
\min_{\substack{v'\in S\setminus \{v\};\\ (w,S\setminus \{v\}, v')\in \mathcal{S}}} \{ \zeta(w,S\setminus \{v\}, v') + c_d(v',v)\} & \text{otherwise}
\end{cases} \label{eq:Bellman}
\end{align}

By construction, for any $w$, $w'$, and $S\subseteq V_d$, the shortest path in $\mathcal{O}$  from the respective initial state $(w, \emptyset, w)$ to terminal state $(w,S,w')$  represents 
an $\mathcal{N_{BS-R}}$-valid operation $o=wsw'$ with $\{s\}=S$ (thus the visiting order) that has the shortest drone flight time, if such an operation exits.   
\subsubsection{Computational complexity of the ops graph for unlimited flight time} \label{sec:ops_complexity}



\begin{proposition} \label{prop:comp_ops}
Consider the BS-R neighbourhood $\mathcal{N}_{BS-R}(x,p)$. 
The computation of the ops graph is polynomial in the input size of the considered DRP-E instance (but exponential in parameter $p$), and requires a worst-case computational time of $O\left(4^{p}\left(n_r^2n_d^2p+n_r n_d^2 p^2\right)\right)$. 
\end{proposition}
\proof
The described dynamic programming approach implied by equations (\ref{eq:Bellman}) has linear time complexity in the number of transition arcs in ops graph $\mathcal{O}$. There are no outgoing arcs from terminal states. Lemma \ref{lem:valid_trans} bounds the number number of transition arcs from each non-terminal state by $n_r+2p$: There are $n_r$ transitions to terminal states and at most $2p-1$ transitions to non-terminal states.
Proposition \ref{prop:comp_ops} from Appendix~ \ref{sec:app_ops_complexity} reports that the total number of non-terminal states in the ops graph is $O(n_d^2 n_r p \cdot 4^p)$. Putting this together, we get the stated complexity. 





\endproof

\subsubsection{Ops graph with energy constraints} \label{sec:feasible_ops}


In case of the limited maximal flight time  $e_{max}$, we prohibit transitions that lead to (energy-)infeasible operations. We use triangular inequalities to perform this in a forward-looking manner.
For each non-terminal state $(w,S,v)\in \mathcal{O}$, we consider the transition to a corresponding 
non-terminal state $(w,S\cup\{v'\},v'=v_i)$ with $v'\in V_d\setminus S$ only if
\begin{align}
\zeta(w,S,v)+c_d(v,v')+\min_{w''\in V_r}\{c_d(v',w'')\}\leq e_{max} \label{eq:energyinterim} 
\end{align}
The (energy-)feasibility check of the transition arcs to terminal states $(w,S,w')$ is straightforward. 






Note that the verification of the energy constraints does not change the time complexity stated in Proposition~\ref{prop:comp_ops}. The distance to the closest RL  $\min_{w\in V_r}\{c_d(v,w)\}$ can be pre-computed for every destination $v\in V_d$ in runtime $O(n_dn_r)$, such that feasibility checks require $O(1)$ for each transition.

\subsection{Meta state graph  }\label{sec:meta}
After we have computed the ops graph, we construct the meta graph to analyse feasible drone tours $\pi_d\in\mathcal{N_{BS-R}}(x,p)$. 
Let \textit{operation set}  $O=wSw', S\subseteq V_d$ denote the set of all feasible, $\mathcal{N}_{BS-R}$-valid operations  $o=wsw'$ with $S=\{s\}$ for given neighborhood $\mathcal{N_{BS-R}}(x,p)$. Let $C_d(O)$ denote the shortest drone flight time over all operations in $O$, if such exist. By construction, the costs $C_d(O), O=wSw',$ correspond to the costs of the terminal states $(w,S,w')$ of the ops graph. 
We use these costs from the ops graph as an input to compute $\mathcal{OP}_{BS-R}^*$ -- the collection of all feasible $\mathcal{N}_{BS-R}$-valid operation sets associated with their corresponding makespan 
$\mathcal{M}(O)=\max \{ c_p(w,w'), C_d(O) \}$ for $O=wSw'$.
Let also denote $\mathcal{RL}$ as the collection of all recharging legs $r\in V_r\times V_r$ associated with their makespans. 
Then the meta graph $\mathcal{G}=\left(\mathcal{V}, \mathcal{A}\right)$ enumerates all sequences of operations in $\mathcal{OP}_{BS-R}^*$ and recharging legs in $\mathcal{RL}$ that form feasible drone tours $\pi^d\in\mathcal{N_{BS-R}}(x,p)$. 

Similarly to the ops graph, the main challenge in the construction of the meta graph is to examine only those states, which are relevant for the construction of BS-R-neighbors of $x$. We call such states $\mathcal{N_{BS-R}}$\textit{-valid} and  limit the state set $\mathcal{V} $ to $\mathcal{N_{BS-R}}$-valid metastates only. In order to efficiently discard $\mathcal{N_{BS-R}}$-\textit{in}valid metastates during the construction procedure, we use a special encoding for the states. This encoding was originally proposed by \citet{balas1999} for the traveling salesman problem (TSP), but we extend it to our problem. 
Before we introduce this encoding, we outline the general architecture of the meta graph $\mathcal{G}$ with an intuitive straightforward description of the states in Section~\ref{sec:meta_general}. In Section~\ref{sec:meta_complexity}, we show how to formulate states in an equivalent way by using the mentioned encoding and state the time complexity of the meta graph. 
\subsubsection{Architecture of the meta graph} \label{sec:meta_general}

\begin{figure}
\centering
\includegraphics[scale=0.5]
{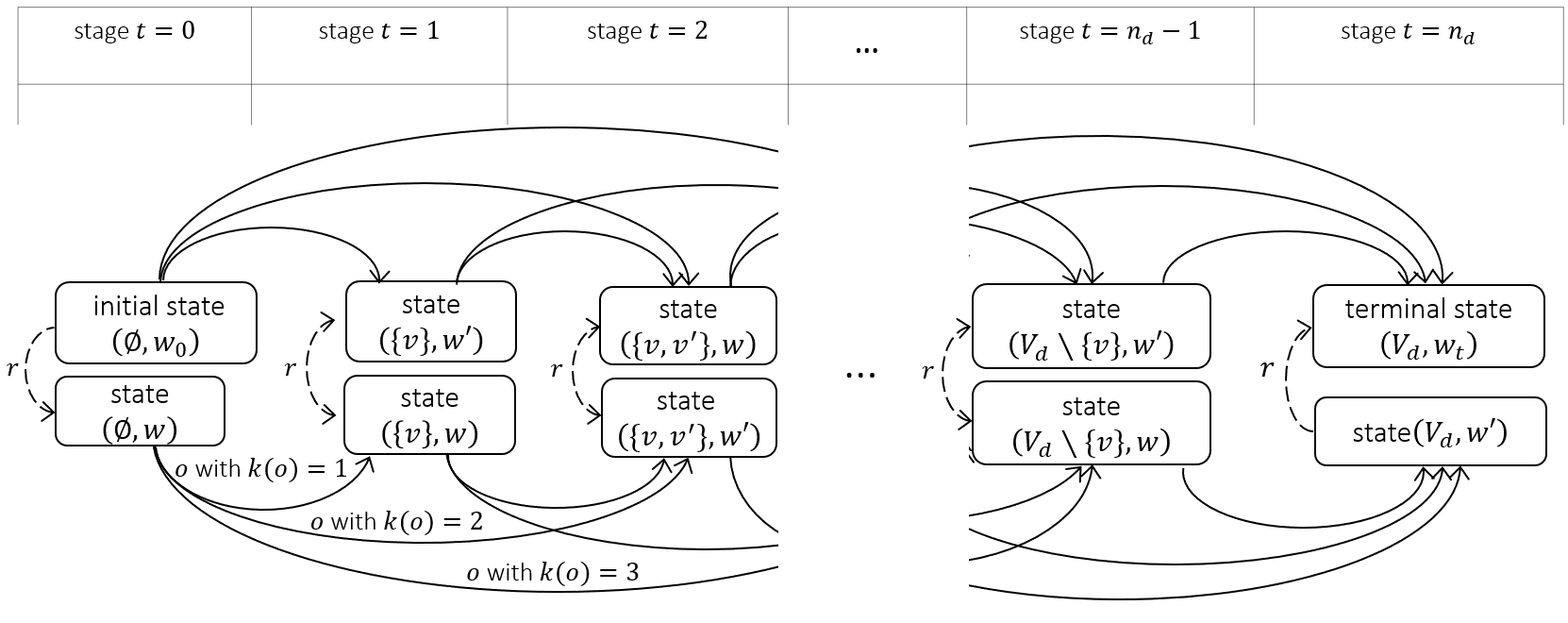}
\caption{A schematic illustration of the meta state graph} \label{fig:metagraph}
\end{figure}

 States in $\mathcal{G}$ are arranged in stages $k \in \{0,1,2,...,n_d\}$. Each state $(S,w)$ at stage $k:=\vert S \vert$ describes the set of visited destinations $S\subseteq V_d$ and the drone's current position at $w\in V_r$. Stage $0$ consists of states  $(\emptyset, w), w\in V_r$, where the \textit{initial state} $(\emptyset, w_0)$ describes the start of each drone tour. At stage $n_d$, we find states  $(V_d,w), w\in V_r$, including \textit{terminal} state $(V_d,w_t)$ that marks the end of each drone tour. 

Transition arcs $\mathcal{A}$ in $\mathcal{G}$ do not necessarily connect metastates at the neighboring stages, but may span over several stages. The overall structure of  $\mathcal{G}$ is depicted in Figure~\ref{fig:metagraph}.  Given state $(S,w) \in \mathcal{V}$, there is a transition to $(T,w') \in \mathcal{V}$  if and only if one of the cases below holds:
\begin{itemize}
    \item $S=T$. Then the transition corresponds to a recharging leg $r=ww'$ and its weight is $\mathcal{M}(r)$. 
    \item $S\subsetneq T$  and the respective operation set $O=wT\setminus Sw'$ is in $\mathcal{OP}_{BS-R}^*$. Then the transition weight is $\mathcal{M}(O)$.
\end{itemize}

By construction, each path in $\mathcal{G}$ corresponds to a feasible drone tour $\pi_d\in \mathcal{N_{BS-R}}(x,p)$, and the shortest of these paths corresponds to the best drone tour $\pi_d^*$ in the considered BS-R neighborhood. We compute the shortest path with the Bellman's equations (\ref{eq:Bellman_meta}).
Recall that it is sufficient to consider  drone tours formed as an alternate sequence of operations and (possibly trivial) two-node recharging legs (see Section~\ref{sec:formal}). This allows us to iterate through the stages of $\mathcal{G}$ in a forward induction manner, combining incoming operation arcs with exactly one two-node recharging leg:
\begin{align}
&    \zeta(S,w)=\begin{cases}
 c_r(w_0,w) & \text{if } S=\emptyset  \\
\min\limits_{w' \in V_r} \{ \epsilon (S,w') + \mathcal{M}(r) \ \vert \ \text{s.t. } r:=w'w \in \mathcal{RL} \} & \text{otherwise}
\end{cases} \label{eq:Bellman_meta}\\
& \text{with } \epsilon (S,w')=  \min\limits_{ \substack{T \subsetneq S;\\ w''\in V_r}}  \{ \zeta(T, w'') + \mathcal{M}(O) \ \vert \ \text{s.t. } (T,w'') \in \mathcal{V} \text{ and }  O =w''\{S\setminus T\}w' \in \mathcal{OP}_{BS-R}^*\} \nonumber
\end{align}


\subsubsection{Alternative representation of states and complexity of the meta graph }\label{sec:meta_complexity}
In the following, we explain the encoding, which enables to construct $\mathcal{N_{BS-R}}$-valid metastates only, so that the meta graph can be computed in polynomial time. The transition arcs and the Bellman's equation remain the same, as stated in Section~\ref{sec:meta_general}. 

The key idea is that we can uniquely describe set $S$ of the $k=|S|$ \textit{first} visited destinations of any BS-R neighbor $\pi_d$ of destination sequence $x$ 
as follows:
\begin{align}
&    S=\{ v_j:  \ j \in ([k]\setminus S^+_k) \cup S^-_k \}, \text{ with} \label{eq:def_encoding}\\
 &   S^-_k:= \lbrace l \in [n_d]: l \geq k+1, v_l \in S \rbrace 
    \text{   and    }
    S^+_k:=  \lbrace h \in [n_d]: h \leq k, v_h \notin S \rbrace \label{eq:encoding}
\end{align}

In other words, $S^-_k$ refers to the destinations that are at positions later than $k$ in $x$ and are moved up (before or at position $k$) in $x'=\pi_d(V_d)$ of neighboring tour $\pi_d$. $S^+_k$ refers to the destinations that are at positions no later than $k$ in $x$ and are moved down (after position $k$) in $x'=\pi_d(V_d)$.

\begin{example}\label{ex:notation}
Consider the drone tour $\pi_d=(w_0,w_2,v_4,v_1,v_5,w_1,w_1,v_2,v_3,w_2,w_t)\in \mathcal{N_{BS-R}}(x,4)$, with $x=(v_1,v_2,v_3,v_4,v_5)$. Then the set of  $k=3$ first visited destinations in $\pi_d$ is $S=\{v_4,v_1,v_5\}$, which can be uniquely described by the index sets $S_3^-=\{4,5\}$, $S_3^+=\{2,3\}$.
\end{example}

With this new notation, we represent metastates as $(S_k^-,S^+_k,w)$, where $k$ refers to the stage number. 
In this encoding,  $\mathcal{N_{BS-R}}$-valid metastates and valid transitions between them have \textit{the same structure} for all the stages, which solely depends on parameter $p$ and not on the instance-specific parameters 
 (see Propositions~\ref{prop:valid_encoding} and \ref{prop:metaarcs}). 
 As the result, these can be computed as a lookup table in the preprocessing, see Table \ref{tab:lookup}.



%

\begin{proposition}\label{prop:valid_encoding}
For the BS-R neighborhood $\mathcal{N_{BS-R}}(x,p)$, the encoded state $(S^-_k, S^+_k,w)$ is a $\mathcal{N_{BS-R}}$-valid metastate at stage $k\in[n_d]$ if and only if the following conditions hold:
\begin{align}
& - \quad |S^-_k|=|S^+_k| \leq \frac{p}{2} \label{eq:k_cardinality} \\
& - \quad S^-_k\subseteq\{k+1, \ldots,k+p-1\} \text{, }S^+_k\subseteq\{k-p+2, \ldots,k\}   \label{eq:k_candidates} \\
& - \quad \max\{l:l\in S^-_k\}-\min\{h:h\in S^+_k \} < p  \label{eq:k_distance}
\end{align} 
\end{proposition}
\proof
See Appendix~\ref{sec:app_meta}. 
\endproof



 \begin{proposition}\label{prop:metaarcs}
 There is an arc between metastate $(S^-_k,S^+_k,w) \in \mathcal{V}$ at stage $k$  and metastate $(S^-_{k+h},S^+_{k+h},w')\in \mathcal{V}$ at stage $k+h, h\geq0$  if and only if one of the cases below holds:
 \begin{itemize}
 \item $h=0$ and $S^-_{k+h}=S^-_k$, $S^+_{k+h}=S^+_k$
 \item $h\geq 1$, operation set $O=wHw'$ that corresponds to the transition arc belongs to $\mathcal{OP}^*_{BS-R}$ and the following conditions hold simultaneously:  
     \begin{align}
     & - \quad \lbrace j \in S_k^-: j> k+h \rbrace \subseteq S^-_{k+h}  \label{eq:metaarc_cond1} \\
     & - \quad \lbrace j\in S^-_k:j\leq k+h \rbrace \cap S^+_{k+h}= \emptyset  \label{eq:metaarc_cond2} \\
     & - \quad \lbrace j\in S^+_{k+h} : j \leq k \rbrace \subseteq S^+_k  \label{eq:metaarc_cond3}
     \end{align}
     Thereby $H=\{v_l \ : \ l\in I\}, \ I= (\lbrace k+1, ..., k+h \rbrace \cup S^+_k \cup S^-_{k+h}) \setminus (S^+_{k+h} \cup S^-_k)$.
 \end{itemize}
 \end{proposition}
\proof
see Appendix~\ref{sec:app_meta}. 
\endproof

The proof of Proposition~\ref{prop:nbr_metastates} in Appendix~\ref{sec:app_meta} sketches the construction procedure for $\mathcal{N_{BS-R}}$-valid metastates at any stage $k$ based on the rules of Proposition \ref{prop:valid_encoding}. For fixed stage $k$, all these metastates can be constructed from a well-specified list of valid combinations $(S^-_k,S^+_k)$. The first and second columns of the lookup table (see Table~\ref{tab:lookup}) provide an example of such list for $p=4$. 
 Metastates can then be computed by replacing the parameter value $k$ of each valid $(S^-_k,S^+_k)$ from this lookup table by the stage number and combining it with each $w\in V^r$.

Similarly, we can quickly identify all valid transitions in the meta graph based on pre-computed lists, which we construct from Conditions~(\ref{eq:metaarc_cond1})-(\ref{eq:metaarc_cond3}) of Proposition \ref{prop:metaarcs} (see the remaining columns of Table~\ref{tab:lookup} for an example).

\begin{example}
Let $x=(v_1,v_2,v_3,v_4,v_5)$ and $p=4$. Then there are 8 valid pairs of $(S^-_3,S^+_3)$ at stage $k=3$, which results in $(8\cdot n_r)$ $\mathcal{N_{BS-R}}$-valid metastates (see Table \ref{tab:lookup}). For example, $(S^-_k,S^+_k)$-pattern no. 8 describes metastates $(S_3^-=\{4,5\},S_3^+=\{2,3\},w), w\in V_r$. The third column of Table \ref{tab:lookup} states valid transitions to stage $k+1=4$. These are transitions to states $(S_4^-=\{5\},S_4^+=\{2\},w'), w'\in V_r$ [$(S^-_k,S^+_k)$-pattern  no. 2] and $(S_4^-=\{5\},S_4^+=\{3\},w'), w'\in V_r$ [$(S^-_k,S^+_k)$-pattern no. 3].
\end{example}




 \begin{table}
\caption{Lookup table for $\mathcal{N_{BS-R}}$-valid transitions for $p=4$ between stage $k$ and stage $k+h$, $h\leq 3$ }  
\label{tab:lookup}
\begin{center}
\scriptsize{\begin{tabular}{p{1.6cm}p{3.5cm}|p{3.1cm}p{3.1cm}p{3.1cm}}
	\toprule
 & Valid pairs of & \multicolumn{3}{c}{Valid transitions} \\   
 No. & $(S^-_k,S^+_k)$ &  stage $k+1$ & stage $k+2$ & stage $k+3$ \\
  \midrule
  1 & $(\emptyset,\emptyset)$ & 1,5,6,7 & 1,3,4,5,6,7,8 & 1,2,3,4,5,6,7,8 \\
  2 & $(\lbrace k+1\},\{k-2\})$ & 1 & 1,5,6,7 & 1,3,4,5,6,7,8, \\
  3 & $(\{k+1\},\{k-1\} )$ & 1,2 & 1,5,6,7 & 1,3,4,5,6,7,8 \\
  4 & $(\{k+2\},\{k-1\})$ & 2,5 & 1,3,4 &  1,2,5,6,7 \\
  5 & $(\{k+1\},\{k\})$ & 1,3,4 & 1,2,5,6,7 & 1,3,4,5,6,7,8 \\
  6 & $(\{k+2 \},\{k\})$ & 3,5,8 & 1,2,3,4 & 1,2,5,6,7 \\
  7 & $(\{k+3\},\{k\})$ & 4,6,8 & 2,3,5,8 & 1,2,3,4 \\
  8 & $(\{k+1, k+2\},\{k,k-1\})$ & 2,3 & 1,2 & 1,5,6,7 \\
 \midrule[\heavyrulewidth]
  \end{tabular}}


\footnotesize{The values of columns 2 and 1 list valid pairs of $(S^-_k,S^+_k)$ and provide their IDs (numbering), respectively. For each valid pair $(S^-_k,S^+_k)$, columns 3-5 list the IDs (cf. column 1) of valid pairs $(S^-_{k+h},S^+_{k+h})$   which can be part of the head states of outgoing valid transition arcs in the meta graph.}
\end{center}
\end{table}

We conclude by stating the computational complexity of the meta graph in Proposition~\ref{prop:comp_meta}.

\begin{proposition} \label{prop:comp_meta}
Consider BS-R neighbourhood $\mathcal{N_{BS-R}}(x,p)$. Given the collection of feasible, $\mathcal{N_{BS-R}}$-valid operations $\mathcal{OP}^\ast_{BS-R}$, the best feasible drone tour $\pi_d^*\in \mathcal{N_{BS-R}}(x,p)$ can be determined with the worst case computational complexity of $O(n_d^2 n_r^2\cdot 4^{p})$. 
\end{proposition}
\proof
The best tour $\pi_d^*\in \mathcal{N_{BS-R}}(x,p)$ is determined by solving the shortest path problem in meta graph $\mathcal{G}$ with the Bellman's equations~(\ref{eq:Bellman_meta}). The complexity of this procedure is linear in the number of transition arcs, which is the highest for $e_{max}\rightarrow\infty$. In this case, all outgoing arcs for each state are exactly those listed in the lookup table (cf. Table~\ref{tab:lookup}). By Proposition \ref{prop:nbr_metastates} from Appendix~\ref{sec:app_meta}, there are $\mathcal{O}(n_dn_r\cdot 2^p)$ $\mathcal{N_{BS-R}}$-valid metastates. 
For increasing $h$, the list of valid transitions for each metastate at stage $k$ towards stage $k+h$, converges to the complete set of valid metastates at stage $k+h$ (see Table \ref{tab:lookup}). From this follows the approximation of the time complexity.  
\endproof


\subsection{Special case of a single depot} \label{sec:outline}

Consider instances with a single depot $w_0=w_t$. The first and the last visited destinations of a good drone tour $\pi_d$ will probably be geographically close to $w_0$ and, thus, close to one another. This implies that  it may be reasonable to visit destinations at terminal and initial positions of sequence $x=\pi_d(V_d)$  within the same operation. But BS-R neighbourhood does not allow this if $n_d\gg(2p+1)$ and if the maximal flight time $e_{max}$ is limited. For example, as follows from Lemma~\ref{lem:valid_trans}, $v_1$ and $v_{n_d}$ may belong to the same operation only if destinations $[v_{1+p},v_{n_d-p}]$ belong to it as well. 

The adaptation of neighbourhood $\mathcal{N_{BS-R}}(x,p)$ to the case of the single depot is not straightforward. We proceed as follows: After having examined $\mathcal{N_{BS-R}}(x,p)$ as described above, we create a small number, $p(p-1)$, of additional permutations of $x$ by shifting one node in $x$ either from one of the $(p-1)$ positions in the beginning to one of the $(p-1)$ positions in the end of sequence $x$, or \textit{vice versa}. For each such permutation, we compute an optimal positioning of the RLs in $O\left(n_d^2n_r^2\right)$. 

The computational complexity of VLSN remains unchanged.

\section{An exact solution approach} \label{sec:exact}
One important aspect of VLSN is its scalability with the parameter $p$: With $p \rightarrow n_d$ it converges to an exact approach.  

A more time-efficient exact solution method for DRP-E can be derived directly from the two-stage dynamic programming approach outlined in Section~\ref{sec:vlns} by dropping the restriction to $\mathcal{N_{BS-R}}$-valid states in both state graphs. The logic behind the architecture of these graphs remains the same. Recall that the computational complexity of the constructed dynamic programming procedures linearly depends on the number of arcs in the ops graph and the meta graph, respectively.

As a result, 
the exact procedure for the ops graph has the worst-case runtime of $O(n_rn_d(n_d+n_r)\cdot 2^{n_d})$: the number of non-terminal states $(w,S,v)$ in the ops graph is $O(n_r n_d \cdot 2^{n_d})$, each having up to $O(n_r)$ arcs to terminal states and $O(n_d)$ arcs to non-terminal states.


The meta graph conserves the overall architecture described in Section~\ref{sec:meta_general}, the encoding is not necessary. 
The number of metastates $(S,w)$ is exactly $n_r \cdot 2^{n_d}$. Each metastate $(S,w)$ has at most $O(n_r \cdot 2^{|S|})$ incoming arcs, these are arcs from metastates $(T,w')$ with $T\subseteq S$. Summing up the transition arcs for all metastates, we receive the time complexity of the meta graph of $O\left(\sum_{(S,w)\in\mathcal{V}}{n_r\cdot 2^{|S|}}\right)=O\left(n_r^2 \cdot 3^{n_d}\right)$.

The overall time complexity of the exact algorithm 
equals $O\left(n_r^2 3^{n_d}+n_d n_r\left(n_d+n_r\right) 2^{n_d}\right)$.
In our computational experiments, this exact algorithm solved instances with up to 32 nodes (incl. $n_d=16$ destinations) within 3 minutes on average and within 20 minutes in the observed worst case.

\section{Computational experiments} \label{sec:computational}

We analyse the performance of VLSN by embedding it into local search (VLSN-LS) and variable neighbourhood descent (VLSN-VND) procedures. First, we compare the performance of these algorithms to the state-of-the-art heuristics --\textit{Route, Transform, Shortest Path (RTS)} of \citet{poikonenandgolden2020} and the \textit{\underline{lim}ited \underline{op}erations size heuristic (LIMOP)} based on the ideas of \citet{bouman2017} -- on a structured randomly generated data set.
Afterwards, we illustrate flexibility of the designed algorithms by solving a reach-on-details real-world case study of a missing persons search in the woods around Lake Occhito in Southern Italy. The case study is based on the European cooperation project SHERPA for disaster relief \citep[cf.][]{marconietal2012}.

RTS optimally inserts RLs into the initial sequence of destinations $x^{\text{TSP}}$
. Initial sequence $x^{\text{TSP}}$  corresponds to the shortest path from $w_0$ to $w_t$ that visits all destinations in $V^d$, received by solving the respective TSP. Observe that RTS, in fact, returns a local optimum of $\mathcal{N_{BS-R}}(x^{\text{TSP}},p=1)$. Thus, RTS can be interpreted as a special case of VLSN-LS with $p=1$.

LIMOP enumerates all feasible drone tours, in which the number of destinations $k(o)$ in each operation $o$ does not exceed some value $klim\in\mathbb{N}$. Following \citet{bouman2017}, LIMOP is implemented as a two-stage dynamic programming procedure.

We outline the data generation and experimental setup in Section~\ref{sec:design}. Section~\ref{sec:performance} compares the performance of the proposed algorithms to the state-of-the-art heuristics and shows that VLSN leads to significant improvements in the solution quality. Since the design of VLSN involves an extended analytical apparatus, the question arises whether similar results could have been achieved with more simple techniques. For instance, by embedding the state-of-the-art route-first-split-second heuristic RTS in a metaheuristic procedure. Section~\ref{sec:analysis} reveals that this is not the case as well as quantifies the trade-off between the solution quality and the runtime for different values of parameters $p$ in VLSN.
Section~\ref{sec:casestudy} reports on the case study.

\subsection{Data generation and experiment design} \label{sec:design}

Although several articles examined variants of DRP-E \citep[cf.][]{luoetal2017,
 poikonenandgolden2020}, to the best of our knowledge, there is no benchmark data set for this problem. In this section, we propose a structured data set that describes a wide range of possible drone and rover characteristics.  

\begin{figure}
    \centering
    \includegraphics[scale=0.7]{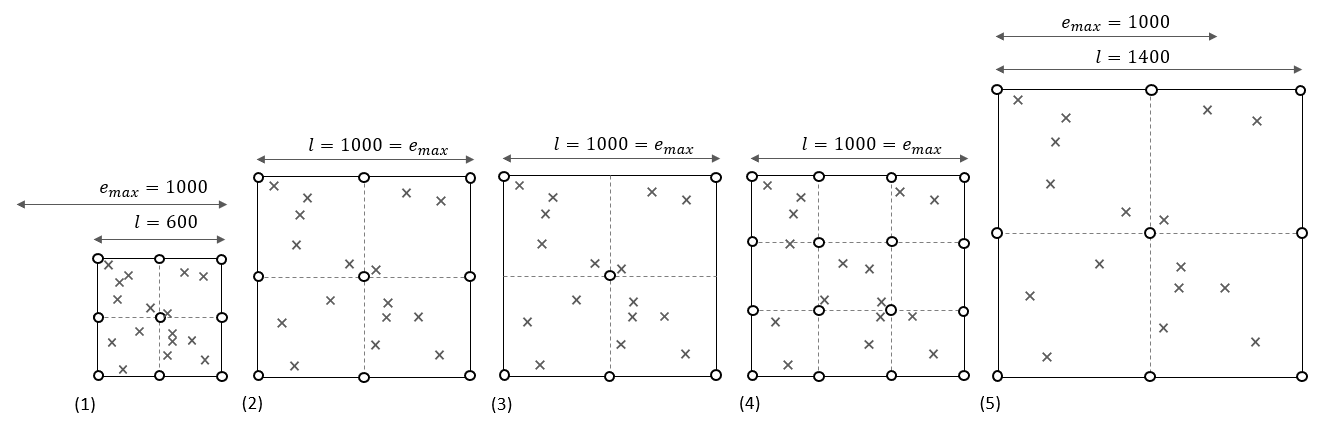}
\caption{Examples of generated instances for Small\\ \footnotesize{(1): DenHigh, (2): Basis, (3): LocLow, (4): LocHigh, (5) DenLow. To keep the proportion of destinations and RLs at 3 in LocLow for Small,  $n_r$ had to be set to a non-quadratic number 6.}} \label{fig:illustration_instances}
\end{figure}

We generate instances by randomly uniformly scattering a given number of destinations $n_d$ in a square $l\times l,l\in\mathbb{N}$. The placement of RLs should ensure the feasibility of instances
. Unfeasible instances occur when at least one destination has no close-by RL, such that the battery life of the drone is insufficient even for a visit in a direct return flight.  Possible workarounds in the literature include dismissal of infeasible instances  \citep[cf.][]{poikonenandgolden2020} or calculation of $e_{max}$ as a function of the realized positions of $v\in V_d$ and $w\in V_r$ \citep{gonzalesetal2020}. For the sake of data generation transparency, and similar to \citet{karakandabdelghany2019}, we decided to place RLs in the nodes of a uniform grid instead (see Figure~\ref{fig:illustration_instances}), having as a consequence that the number of RLs should possibly be \textit{quadratic}. A randomly selected RL is set to be both the initial and the target depot. We use the Euclidean metric to measure the distances flown by the drone and the Manhattan metric to measure the distances traveled by the rover. 

We generate two data sets: 
\begin{itemize}
\item Small data set (\textit{Small}) contains instances of moderate size with $n_d:=16$ destinations and $n_r\in\{6,9,16\}$ RLs.
\item Large data set (\textit{Large}) contains instances with $n_d:=100$ destinations and $n_r\in\{36,49,100\}$ RLs.
\end{itemize}

All the instances within Small were solved to optimality with the exact algorithm described in Section~\ref{sec:exact}. Each data set contains 9 settings with 10 instances each, making 180 instances in total. 

We specify instance settings having in mind the drone speed of 10m/sec with the maximal flight time of about 30 minutes and the rover moving at the rate of 5m/sec. 
Since the ratio of the vehicle speeds and not their absolute values are essential to the performance of solution algorithms, we normalize the drone speed to 1 and condense further instance characteristics to a few essential details. Such normalization makes our results more informative in view of ever changing characteristics of the young drone technology. For example, if we set in our basic setting, called Basis (cf. Table~\ref{tab:instances}), the \textit{time unit} $TU:= 2$ seconds and the \textit{length unit} $LU:= 20$ m, we receive vehicle speeds of 10m/sec and 5m/sec as considered by \citet{poikonenandgolden2020}, combined with the realistic maximal drone flight time $e_{max}=33$ minutes  \citep[cf.][]{Stolaroff.2018}.    

We control the following factors and use one-factor-at-a-time design around the basic settings (marked in bold), see also Table~\ref{tab:OverSettings}:
\begin{itemize}
\item \textit{Rover} speed $\delta\in \{1,\bm{\frac{1}{2}},\frac{1}{3}\}$ \textit{TU/LU}
. The lower is $\delta$, the larger is the impact of the waiting times for the rover on the routing decisions of the drone. 

\item \textit{Energy capacity} $e_{max}\in \{750, \bm{1000}, 1250\}$ \textit{TU}. 

\item \textit{The ratio of the number of destinations and RLs} $\frac{n_d}{n_r}\approx 1,2$ or $3$. Given $n^d=16$ for Small and $n_d=100$ for Large, we set $n_r\in \{6,\bm{9},16\}$ for Small and $n_r\in \{36,\bm{49},100\}$ for Large to achieve these values.  

\item \textit{The number of destinations per square unit (density)} $d  \in \{8,16,45\}\times 10^{-6}$ $LU^{-2}$. To achieve these values, we set $l\in \{600,\bm{1000},1400\}$ LU for Small and $l\in \{1500,\bm{2500},3500\}$ LU for Large.

\end{itemize}

\begin{table}[htp]
\caption[Table caption text]{Overview over the nine settings in each data set -- Small and Large }  \label{tab:instances}
\label{tab:OverSettings}
\begin{center}
\scriptsize{
\begin{tabular}{lcccc}
	\toprule
    
  Setting & ~~Replenishment station speed  $\delta$~~ & ~~Energy capacity  $e_{max}$~~  & ~~Location ratio  $\frac{n_d}{n_r}$~~ & ~~Density  $d$ $(\times 10^{-6})$~~ \\
  \midrule
  Basis & $^1/_2$ & 1000 & 2 & 16 \\

  SpLow & $\bm{^1/_3}$ & \textcolor{gray}{1000} & \textcolor{gray}{2} & \textcolor{gray}{16} \\

  SpHigh  & \textbf{1} & \textcolor{gray}{1000} & \textcolor{gray}{2} & \textcolor{gray}{16} \\ 

  EnLow & \textcolor{gray}{$^1/_2$} & \textbf{\textcolor{white}{0}750} & \textcolor{gray}{2} & \textcolor{gray}{16} \\

  EnHigh & \textcolor{gray}{$^1/_2$} & \textbf{1250} & \textcolor{gray}{2} & \textcolor{gray}{16} \\

  LocLow & \textcolor{gray}{$^1/_2$} & \textcolor{gray}{1000} & \textbf{3} & \textcolor{gray}{16} \\

  LocHigh & \textcolor{gray}{$^1/_2$} & \textcolor{gray}{1000} & \textbf{1} & \textcolor{gray}{16} \\

  DenLow & \textcolor{gray}{$^1/_2$} & \textcolor{gray}{1000} & \textcolor{gray}{2} & \textbf{\textcolor{white}{0}8} \\

  DenHigh & \textcolor{gray}{$^1/_2$} & \textcolor{gray}{1000} & \textcolor{gray}{2} & \textbf{45} \\
  \midrule[\heavyrulewidth]
  \end{tabular}
}
\end{center}
\end{table}

The generated instances can be downloaded from the electronic companion to the paper.

We used a laptop with Intel i7-8565U, 1.80 GHz, 32GB RAM and 
implemented the algorithms in Python 3.7. 
For each instance class, we report \textit{average gap} and \textit{worst gap}, which are the observed average and worst relative deviation from optimality (in case of Small) or from the best known solution (in case of Large), respectively.  We abbreviate \textit{percentage points} with pp.
Unless stated otherwise, we set $p=4$ for the BS-R neighbourhood of VLSN, which, according to our preliminary computational tests, provides the best compromise between solution quality and required runtime.

\subsection{Comparative performance of the proposed solution algorithms} \label{sec:performance}

\begin{table}[htp]
\caption[Table caption text]{Algorithms' performance on \textbf{Small} }  \label{tab:exp1}
\begin{center}
\scriptsize{
\begin{tabular}{l|rrrr|rrrr|rrrr}
\toprule
& \multicolumn{4}{c|}{Average gap ($\%$)}&\multicolumn{4}{c|}{Worst gap ($\%$)}&\multicolumn{4}{c}{\# optimum found} \\
Setting  & \multicolumn{1}{c}{VLSN-LS} & \multicolumn{1}{c}{VLSN-VND} & \multicolumn{1}{c}{RTS}  & \multicolumn{1}{c|}{LIMOP}  & \multicolumn{1}{c}{VLSN-LS} & \multicolumn{1}{c}{VLSN-VND} & \multicolumn{1}{c}{RTS}  & \multicolumn{1}{c|}{LIMOP} &  \multicolumn{1}{c}{VLSN-LS} & \multicolumn{1}{c}{VLSN-VND} & \multicolumn{1}{c}{RTS}  & \multicolumn{1}{c}{LIMOP}  \\
\midrule
Basis & 2.12 & \textbf{0.32} & 5.24 &  12.15 & 8.28 & \textbf{3.23}  & 10.10 & 28.78 & 7  & \textbf{9} & 2 & 0 \\

SpLow &   0.29 & \textbf{0.00} & 2.89 & 4.55 & 2.88 & \textbf{0.00}  & 11.09 & 10.69 & 9  & \textbf{10} & 4 & 1 \\

SpHigh &   0.66 & \textbf{0.09} & 2.17 &  20.65 & 4.33 & \textbf{0.89} &  6.98 & 33.92 & 6 &  \textbf{8} & 2 & 0  \\

EnLow &    4.57 & \textbf{2.76} & 16.27 &  3.65 & 15.8 & 15.8 & 41.15 & \textbf{14.81} & 3 & \textbf{6} & 1 & 3 \\

EnHigh &  0.79 & \textbf{0.21} & 6.26 &  25.57 & 2.67 & \textbf{1.02}  &  23.48 & 43.77 & 2 &  \textbf{7} & 0 & 0 \\

LocLow &  \textbf{0.71} & \textbf{0.71} & 2.72 &  12.61 & \textbf{4.75} & \textbf{4.75} &  7.44 & 16.8 & \textbf{8}  & \textbf{8} & 3 & 0 \\

LocHigh &    0.45 & \textbf{0.40} & 1.73 &  5.06 & \textbf{2.82} & \textbf{2.82} &  6.01 & 10.00 & 5 &  \textbf{6} & 2 & 0 \\

DenLow &  4.03 & \textbf{0.00} & 19.11 &  3.58 & 20.99 & \textbf{0.02} &  36.85 & 6.94 & 2  & \textbf{8} & 1 & 0 \\

DenHigh &    1.05 & \textbf{0.33} & 3.92 &  29.38  & 4.46 &  \textbf{2.36} & 9.09 & 37.46 &  6 & \textbf{7} & 3 & 0 \\
\midrule
Total &    1.63 & \textbf{0.54} & 6.70  & 13.02 & 20.99 & \textbf{15.8} & 41.15 & 43.77 & 48 & \textbf{69} & 18 & 4 \\
\midrule[\heavyrulewidth]
  \end{tabular}
}
\end{center}
\end{table}


\begin{table}[htp] 

\caption[Table caption text]{Algorithms' runtimes on \textbf{Small} }  \label{tab:runtimes_exp1}
\begin{center}
\scriptsize{
\begin{tabular}{l|lccll}
\toprule
& \multicolumn{5}{c}{ $\varnothing$ runtime (sec)}\\
Setting  & EXACT & VLSN-LS & VLSN-VND & RTS  & LIMOP  \\
\midrule
Basis & 78.25 & 1.59 & 23.75 &  0.04 & 48.73  \\

SpLow & 77.16   & 1.62 & 20.77 & 0.11 & 47.71  \\

SpHigh &   77.28 & 1.47 & 21.44 &  0.04 & 47.61   \\

EnLow &  46.30 & 1.10 & 17.66 &  0.04 & 27.96 \\

EnHigh &  208.01 & 3.93 & 48.30 &  0.04 & 110.28 \\

LocLow &  37.74 & 1.03 & 12.09 &  0.03 & 24.60  \\

LocHigh &    339.36 & 6.84 & 71.59 &  0.04 & 208.22  \\

DenLow &  37.67 & 0.08 & 12.39 &  0.04 & 24.47  \\

DenHigh &  972.19   & 4.76 & 108.07 &  0.04  & 231.49  \\
\midrule
Total &   208.22 & 2.57 & 33.60  & 0.05 & 85.67 \\
 \midrule[\heavyrulewidth]

  \end{tabular}}
  \end{center}

\end{table}


\begin{table}[htp] 
\caption[Table caption text]{Algorithms' performance on \textbf{Large}}  \label{tab:exp2}
\begin{center}
\scriptsize{
\begin{tabular}{l|lll|lll|lll|lll}
\toprule
& \multicolumn{3}{c|}{Average gap ($\%$)}&\multicolumn{3}{c|}{Worst gap ($\%$)}& \multicolumn{3}{c|}{\# best found} & \multicolumn{3}{c}{$\varnothing$ runtime (sec)} \\
Setting  & \multicolumn{1}{c}{VLSN} & \multicolumn{1}{c}{VLSN} & \multicolumn{1}{c|}{RTS}   & \multicolumn{1}{c}{VLSN} & \multicolumn{1}{c}{VLSN} & \multicolumn{1}{c|}{RTS}  &  \multicolumn{1}{c}{VLSN} & \multicolumn{1}{c}{VLSN} & \multicolumn{1}{c|}{RTS} &  \multicolumn{1}{c}{VLSN} & \multicolumn{1}{c}{VLSN} & \multicolumn{1}{c}{RTS}     \\
  & \multicolumn{1}{c}{-LS} & \multicolumn{1}{c}{-VND} &    & \multicolumn{1}{c}{-LS} & \multicolumn{1}{c}{-VND} &   &  \multicolumn{1}{c}{-LS} & \multicolumn{1}{c}{-VND} &  &  \multicolumn{1}{c}{-LS} & \multicolumn{1}{c}{-VND} &      \\
\midrule
Basis & 1.36 & 0.00 & 5.12 & 4.46 & 0.00 & 8.87 & 0 & 10 & 0 & 179 & 1800 & 10\\

SpLow &   1.16 & 0.00 & 6.17 & 2.47 & 0.00 & 10.43 & 0 & 10 & 0 & 193 & 1800 & 7\\

SpHigh &   0.72 & 0.00 & 2.37 & 2.93 & 0.00 & 4.46 & 2 & 10 & 0 & 173 & 1800 & 6\\

EnLow &    2.02 & 0.00 & 7.35 & 4.31 & 0.00 & 12.62 & 2 & 10 & 0 & 99 & 1800 & 6\\

EnHigh &  0.63 & 0.00 & 2.14 & 2.39 & 0.00 & 3.85 &  4 & 10 & 0 & 269 & 1800 & 8\\

LocLow &  2.42 & 0.00 & 7.68 & 5.22 &  0.00 & 14.07 & 1 & 10 & 0 & 83 & 1800 & 7\\

LocHigh &    0.23 & 0.00 & 1.29 & 0.85 & 0.00 & 2.49 & 6 & 10 & 0 & 499 & 1800 & 8\\

DenLow &  1.99 & 0.00 & 7.81 & 3.37 & 0.00 & 10.43 & 0 & 10 & 0 & 92 & 1800 & 6\\

DenHigh &    0.35 & 0.00 & 1.58 & 2.84 & 0.00 & 3.95 & 5 & 10 & 0 & 544 & 1800 & 7\\
\midrule
Total &    1.21 & 0.00 & 4.61 & 5.22 & 0.00  & 14.07 & 20 & 90 & 0 & 237 & 1800 & 7\\
\midrule[\heavyrulewidth] 
  \end{tabular}
}
\end{center}
\end{table}

Tables~\ref{tab:exp1}, \ref{tab:runtimes_exp1} and \ref{tab:exp2} report the performance of the heuristic algorithms on Small and Large. We initialize $p_0=2$ for VLSN-VND and set $p\leq 8$ as stopping criterion on Small as well as limit its runtime to 1800 sec on Large. Due to the long runtimes, we set $klim$ to 2 in LIMOP on Small and had to omit LIMOP from the experiments on Large, since it output no results within 1800 sec.

VLSN-LS and VLSN-VND found optimal solutions for 53\% (i.e., 48 out of 90) and 77\% (i.e., 69 out of 90) of instances in Small, respectively. The absolute improvement of VLSN-VND solutions over those of RTS and LIMOP amounted to 6 pp and 12.5 pp on average on Small. The corresponding improvements of VLSN-LS equaled 5 pp and 11.5 pp on average, respectively. In relative terms, VLSN-VND reduced the gap to optimality over RTS for Small by impressive $(6.70-0.54)/6.70=92\%$. For Large instances, VLSN-VND (resp. VLSN-LS) improved the results of RTS by about 4.5 pp (resp. 3.5 pp) on average. Such levels of makespan improvement are of practical significance for many commercial applications. For some instances, the improvement over RTS reached impressive 37 pp in Small and 14 pp in Large in case of VLSN-VND. Contrary to RTS, the performance quality of VLSN-LS and VLSN-VND remains high for a wide range of instances. 

As expected, the gap in performance between VLSN-LS (resp. VLSN-VND) on one hand and RTS on the other is larger in the settings, in which the synchronization of the drone and rover routes is harder to achieve, meaning that the initial TSP sequence of destination visits may turn out to be less attractive. These are EnLow (low energy capacity of the drone) and DenLow (larger average distances between destinations), in which many energy replenishment stops are required. Note, however, that these are also settings SpLow (low speed of the rover) and LocLow (low number of RLs per destination), in which each required replenishment stop gets more expensive in terms of the increased makespan. On the other hand, the relative advantage over the LIMOP gets larger on the EnHigh and DenHigh settings, in which longer drone operations are possible.

VLSN-LS took 2.57 sec on average for instances in Small and about 4 min on average for instances in Large. Computational times of VLSN-VND equaled 33.6 sec and 30 minutes, respectively. 

\subsection{VLSN vs. simpler metaheuristic techniques} \label{sec:analysis}

\begin{table}[htp] 
\caption[Table caption text]{The number of destinations' sequences examined by VLSN}    \label{tab:enumeration}
\begin{center}
\scriptsize{
\begin{tabular}{l|rrrr}
\toprule
& \multicolumn{4}{c}{\# of destinations } \\
Parameter $p$  & \multicolumn{1}{c}{$n^d=5$} & \multicolumn{1}{c}{$n^d=7$} & \multicolumn{1}{c}{$n^d=9$}   & \multicolumn{1}{c}{$n^d=11$}     \\
\midrule
2 & 10 & 23 & 57 & 146 \\

3 &   31 & 130 & 594 & 2,807 \\

4 &   62 & 411 & 3144 & 22,728\\


\midrule[\heavyrulewidth] 
  \end{tabular}
}
\end{center}
\end{table}

As discussed in the introduction to Section~\ref{sec:computational}, VLSN dominates RTS, but it also requires larger runtimes. The additional time is nontrivially used to  explore exponentially many promising solutions in an efficient manner. In the following, we quantify the number of feasible solutions examined by VLSN as well as achieved solution qualities and runtimes at different values of $p$. Afterward, we explore whether similar in quality results could have been achieved with straightforward techniques, such as by embedding the fast state-of-the-art heuristic RTS in basic metaheuristic procedures.

Recall that VLSN, as opposed to VLSN-LS and VLSN-VND, refers to the search of the \textit{(one)} neighborhood of the initial sequence $x$.

Table~\ref{tab:enumeration} exactly enumerates sequences of destinations contained in one BS-R neighborhood of some $x$ for small values of $p$ and $n_d$ assuming unlimited flight time $e_{max}$. Already for $n_d=11$ and $p=4$, VLSN examines more than 22,000 sequences of destination visits. Observe that for each such sequence, we insert RLs optimally. This is $>22,000$ times more than RTS, which examines only one sequence of destination visits. Recall that VLSN-LS, which potentially examines several neighborhoods until the local optimum is found, requires only about $\frac{2.57}{0.05}=51$ times more time than RTS for even larger instances with $n^d=16$ (see Section~\ref{sec:performance}). So that, very conservatively, VLSN needs at least $\frac{22,000}{51}\approx 430$ times less time to examine one feasible sequence of destinations and to find the best replenishment scheme for it than RTS. 

\begin{table}[htp] 

\caption[Table caption text]{Performance dynamics of VLSN, \textbf{Basis setting of Small}}  \label{tab:panalysis}
\begin{center}
\scriptsize{
\begin{tabular}{l|llllllllll}
\toprule
&\multicolumn{10}{c}{Parameter $p=$}\\
 & 1 & 2 & 3 & 4 & 5 & 6 & 7 & 8 & 9 &10\\
\midrule 
Avg gap, \% & 5.4 & 4.1 & 3.4 & 2.2 & 1.3 & 1.3 & 1.2 & 0.6 & 0.3 & 0.3 \\
Runtime, sec & 0.18 & 0.29 & 0.56 & 1.10 & 2.07 & 3.66 & 7.04 & 14.71 & 34.86 & 89.12 \\
\midrule[\heavyrulewidth]

\end{tabular}
}
\end{center}
\end{table}

\begin{table}[htp] 
\caption[Table caption text]{Performance of simpler metaheuristics compared to VLSN on \textbf{Large} within equal runtimes}  \label{tab:RTS_meta}
\begin{center}
\scriptsize{
\begin{tabular}{l|cc|ccc|ccc}
\toprule
& \multicolumn{2}{c|}{Average \# of examined destinations' sequences} & \multicolumn{3}{c|}{\# RTS-improvements} & \multicolumn{3}{c}{$\varnothing$ RTS-improvement ($\%$)} \\
Setting   &  \multicolumn{1}{c}{RTS-3NN}  &  \multicolumn{1}{c|}{SA-RTS-3OPT} & \multicolumn{1}{c}{VLSN} & \multicolumn{1}{c}{RTS-3NN} &  \multicolumn{1}{c|}{SA-RTS-3OPT} & \multicolumn{1}{c}{VLSN} & \multicolumn{1}{c}{RTS-3NN} & \multicolumn{1}{c}{SA-RTS-3OPT}    \\
\midrule
Basis & 211 & 169  & 10 & 0 & 2 & 3.30 & 0.00 & 0.11\\

SpLow &   221 & 127  & 10 & 0 & 3 & 4.00 & 0.00 & 0.09\\

SpHigh &  219 & 152  & 10 & 0 & 1 & 1.47 & 0.00 & 0.01\\

EnLow &   114 & \textcolor{white}{0}72 & 10 & 0 & 4 & 4.09 & 0.00 & 0.53\\

EnHigh & 291 & 158  &  10 & 0 & 1 & 1.32 & 0.00 & 0.00\\

LocLow & 116 & \textcolor{white}{0}31  & 10 & 0 & 0 & 3.79 & 0.00 & 0.00\\

LocHigh &   463 & 307 & 10 & 0 & 2 & 1.03 & 0.00 & 0.01\\

DenLow &   108 & \textcolor{white}{0}52  & 10 & 0 & 0 & 4.33 & 0.00 & 0.00\\

DenHigh &    434 & 374  & 10 & 0 & 3 & 0.84 & 0.00 & 0.01\\
\midrule
Total &     242 & 160 &  90 & 0 & 16 & 2.68 & 0.00 & 0.08 \\
\midrule[\heavyrulewidth] 
  \end{tabular}
}
\end{center}
\end{table}

Table~\ref{tab:panalysis}
reports the trade-off between the computational time and the solution quality of VLSN with different values of $p$. We depicted only the Basis setting of Small, as the results of the remaining settings are similar. With each increase of $p$ by 1, the runtime about doubles on average.

 Finally, Table~\ref{tab:RTS_meta} compares VLSN with $p=4$ to simpler (meta-) heuristics on Large for the same runtime (VLSN runtime): Those are the repeated application of RTS (\textit{RTS-3NN}), which restarts RTS at each iteration with a randomized 3-nearest neighbour, and RTS embedded into the simulated annealing with a 3-exchange of the arcs on the destination sequence as perturbation (\textit{SA-RTS-3OPT}). 
RTS-3NN (resp. SA-RTS-3OPT) performed \textit{at most} 573 iterations (resp. 734 iterations) per instance. In other words, just a fraction of destination sequences have been examined in the same runtime compared to VLSN (cf. Table~\ref{tab:enumeration}). 
As a consequence, RTS-3NN did not manage to improve any of the (single-run) RTS solutions, and SA-RTS-3OPT improved the initial RTS solutions only marginally in 16 of 90 tested instances. In contrast, a single BS-R neighbourhood search with VLSN outperformed the state-of-the-art RTS solution in all tested instances with an average improvement of 2.68 \%.


\subsection{Case study } \label{sec:casestudy} 

In this section, we demonstrate the flexibility of the proposed VLSN-LS approach by testing it in a real-life search operation environment. This case study extends the basic formulation of DRP-E by a number of realistic and technical problem features. VLSN-LS can easily accommodate those features without a change in its general structure. We benchmark the generated solution against the state-of-the-art approach currently used by practitioners in the field.

\begin{figure}
\centering
\includegraphics[width = 0.5\columnwidth]{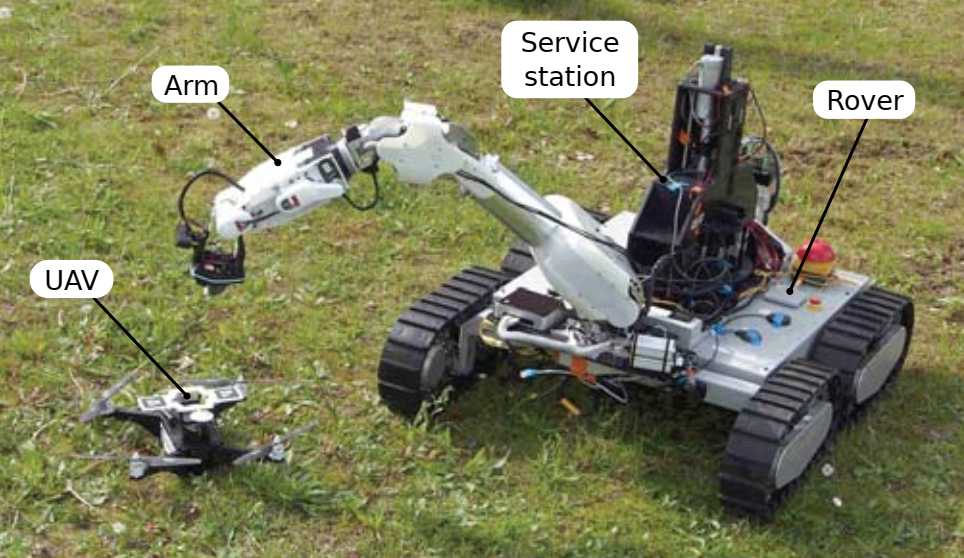}
\caption{The robotic system designed and tested in \cite{Barrett.2018} \\ \footnotesize{The rover is equipped with a robotic arm able to grasp, manipulate and release the drone. Moreover, the rover is equipped with a service station that automatically swaps the discharged drone battery with a fully charged one.}} \label{fig:SherpaRobots} 

\end{figure}

Searching for missing people in the woods and mountains is an important area of application for drones since they can be deployed in the shortest time, at a low cost and can be utilized in dangerous and otherwise hard to access regions. In U.S. national parks alone, about 11 Search and Rescue (S\&R) operations take place each day \citep{Heggie.2009}. Drones use manifold technologies that enable searching even at night and in densely vegetated area, including thermal imaging cameras and sensors for locating mobile phone signals.

In our case study, we focus on a S\&R mission for a lost mushroom hunter in a mountainous area  \citep[see][for more details]{marconietal2012}. Our case study features the deployment of a small-scale quadrotor drone and a mobile replenishment station mounted on a ground rover (see Figure \ref{fig:SherpaRobots}), developed for S\&R missions in the European project SHERPA \citep{marconietal2012}. The search region for missing persons of $\approx 7$ km$^2$ covers the forest area of Occhito, in Southern Italy. The area has to be examined in the quickest possible way in order to intercept any potential life threatening situations that a missing person may be facing. 

\begin{figure}[!tbp] 
  \begin{subfigure}[b]{0.495\textwidth}
    \includegraphics[width=\textwidth]{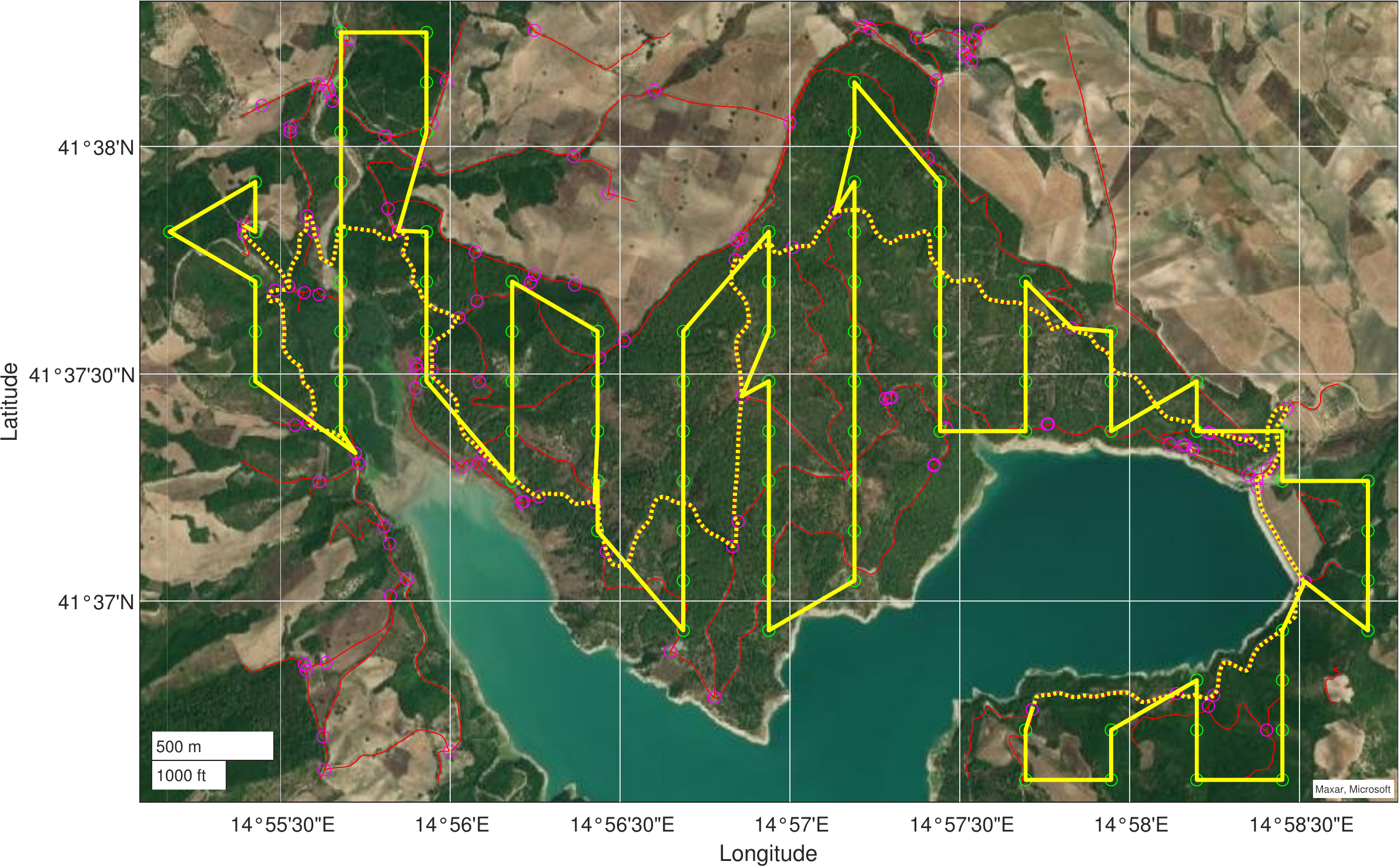}
    \caption{\footnotesize{Tour generated by PRACT}}
    \label{fig:case_study}
  \end{subfigure}
  \hfill
  \begin{subfigure}[b]{0.495\textwidth}
    \includegraphics[width=\textwidth]{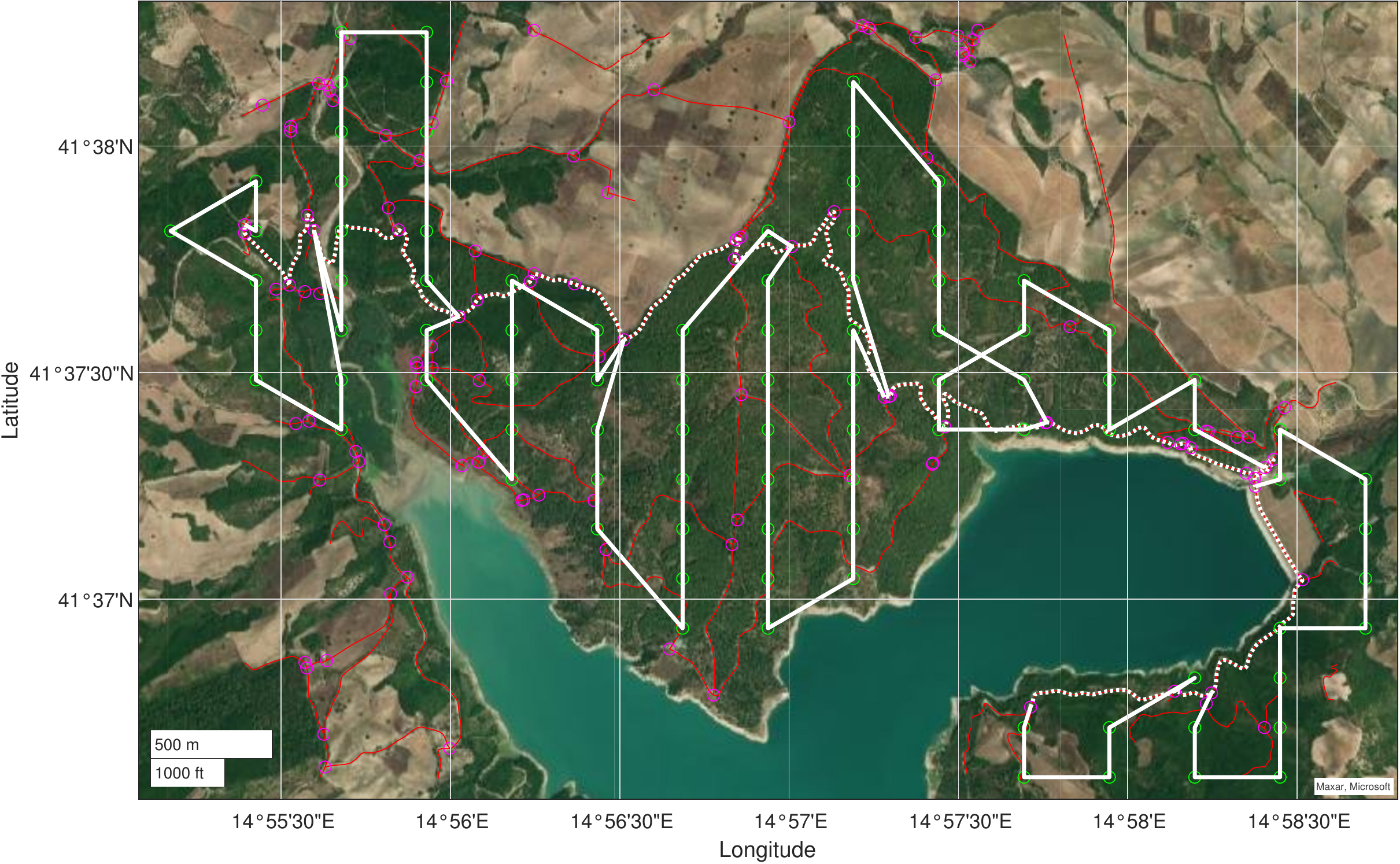}
    \caption{\footnotesize{Tour generated by VLSN-LS}}
    \label{fig:f2}
  \end{subfigure}
  \caption{ The results of the case study -- a typical summer S\&R scenario in mountains \\ \footnotesize{Red lines mark ground trails, magenta and green circles denote RLs and destinations, respectively. Dotted lines show the rover path and continuous lines depict the drone route of the respective solution.}}   \label{fig:fafb} 
\end{figure}

Along the walking trails of the woods, 115 potential RLs are designated as meeting-points for the vehicles, and 100 destinations are identified for the drone to cover the search area (see Figure \ref{fig:fafb}). The rover is assumed to travel at a constant speed of $1$ m/s, which represents a practical limitation due to the roughness of mountain terrain. The drone flies at a constant altitude of $100$ m above the highest point of the intended area and at constant speed of $3$ m/s with an approximate energy consumption rate of $r_{fl} = 15806$ mA.
(see Appendix~\ref{sec:AppDrone} for details on the calculation).

Each landing of the drone on the rover and each take-off of the drone from the rover requires additional time -- $c_{land}=67$ s and $c_{tkof}=33$ s, respectively -- and  consumes additional energy of $\xi_{land}\approx15202$ mAs and $\xi_{tkof}\approx581658$ mAs, respectively. As described in \cite{Barrett.2018}, an automated battery swap, including the time required for locating the drone, grasping and loading it on the service station, changing the battery, unloading the drone from the service station, and releasing the drone on ground amounts to $c_{swap}=250 s$. The drone's battery has the total capacity of $\xi_{max}=21.6\,e6$ mAs, corresponding to $6$ Ah, which is the typical value of 4-cells LiPo batteries. To cover for variations in altitude, uncertain events like wind and the non-linear energy consumption at the end of the battery life, we introduce a residual energy of $10 \%$ of $\xi_{max}$ which the drone cannot fall below. The risk of theft and vandalism increases if the drone waits unattended for the rover on the ground. Therefore, the drone can only wait while hovering at the constant altitude of 100m above the RL, which corresponds to the energy consumption rate $r_{hov}=15687$ mA.

All specifics of the case study can be easily incorporated into the developed VLSN-LS. Figure~\ref{fig:fafb} depicts the vehicle routes suggested by VLSN-LS with 
$p=4$ and 
the state-of-the-art heuristic (PRACT) developed 
in the context of the EU Project SHERPA
(see Appendix \ref{sec:app_pract} for more details). 

\begin{table}[htp]
\caption{Comparison of the solutions by VLSN-LS and the state-of-the-art in practice PRACT }  \label{tab:case_study}
\begin{center}
\scriptsize{

\begin{tabular}{l|rr}
\toprule
  & \multicolumn{1}{c}{VLSN-LS} & \multicolumn{1}{c}{PRACT}   \\
\midrule
\# operations & 9 & 8 \\
Avg mutual waiting time, sec & 82 & 631 \\
Max mutual waiting time, sec & 347 & 1,113 \\
Avg battery utilization, $\%$  & 81 & 82\\
Max battery utilization, $\%$ & 90 & 90\\
Runtime, sec & 1854.6 & 0.2 \\
\textbf{Makespan of the search operation, sec} & \textbf{13,040} & \textbf{16,244} \\
\midrule[\heavyrulewidth]
  \end{tabular}
}
\end{center}
\end{table}

In about half an hour of computing time, VLSN-LS improved the state-of-the-art from practice by 20\%: from a 4 h 31 min makespan of the PRACT's solution down to 3 h 37 min (see Table~\ref{tab:case_study}). The algorithms differ in their ability to handle the synchronization of the rover and the drone. PRACT aspires for routes with less and longer drone operations, without accounting for the longer travel times of the slow rover. The drone's maximum waiting time for the rover equaled 19 min in the PRACT's solution of this case study.  VLSN-LS, in contrast, uses shorter operations to synchronize the drone and rover routes, such that the maximum mutual waiting time was less than 6 min. Furthermore, PRACT does not construct recharging legs, whereas VLSN-LS' solution contains three of them. Finally, PRACT cannot guarantee that the remaining drone's energy suffices for hovering, when waiting for the rover is required. In the current benchmark solution, the drone was forced to land and switch off several times due to the depletion of energy.

\section{Conclusions} \label{sec:conclusion}
In this paper, we addressed a basic problem that is relevant for many drone applications: the drone has to visit several destinations and at some points in-between, at either a stationary or mobile replenishment station, swaps (or replenishes) its battery. We call this basic problem, the drone routing problem with energy replenishment (DRP-E). To solve DRP-E, we propose a non-trivial very large-scale neighborhood search VLSN. VLSN synergetically leverages two large-sized polynomialy solvable DRP-E \underline{s}ub\underline{p}roblems (SP1 and SP2). Roughly speaking, the number of examined feasible solutions by VLSN is a multiple of those in
SP1 and SP2, and, thus, grows exponentially in the size of the DRP-instance, whereas the computational time remains polynomial. 

The size of the associated neighborhood and, thus, the computational complexity of VLSN can be flexibly controlled via parameter $p$. If $p=1$, the solutions of VLSN are equivalent to the state-of-the-art algorithm RTS \citep{poikonenandgolden2020}, and  with large enough values of $p$, VLSN turns into an exact algorithm, which is the first exact algorithm proposed for DRP-E. 

In our computational experiments, algorithms based on VLSN -- local search VLSN-LS and variable neighborhood descent VLSN-VND -- outperformed state-of-the art heuristics by a significant margin. This result is important, especially since the state-of-the-art heuristic of \citet{poikonenandgolden2020} is notoriously hard to improve, at least by embedding it into basic metaheuristic procedures (see Section~\ref{sec:analysis}). The exact algorithm solved instances with up to 32 nodes (of it 16 destinations) within 3 minutes on average and within 20 minutes in the observed worst case. This is a very competitive result, e.g., compared to the current state-of-the-art by \citet{Roberti.2021} for the similar \textit{TSP with Drone} problem, which solves instances with 39 nodes to optimality. 
The benchmark data set proposed in this article can be downloaded from the electronic companion.

In our case study, we extended the general formulation of DRP-E with a number of realistic operational and technical features for a search and rescue scenario in a wooded area of Southern Italy. VLSN-LS easily accommodated these additional features and improved the state-of-the-art solution of practitioners by 20 $\%$, reducing the search operation by 54 minutes.

  
Future research should address how to extend the proposed VLSN methodology to a number of non-trivial extensions of DRP-E, such as multiple drones or rovers, stochastic flight times, or partial battery replenishment. Important application-specific features also include complex energy-consumption functions, data transmission ranges or moving persons in search and rescue operations. 

\textbf{Acknowledgment}

This project was partially funded by the Deutsche Forschungsgemeinschaft (DFG - German Research Foundation) as part of the SuPerPlan project GZ: OT 500/4-1.


\bibliographystyle{unsrtnat}
\bibliography{dronesAO}
%
\newpage
\setcounter{page}{0}
    \pagenumbering{Roman}
    \setcounter{page}{1}
\renewcommand{\thesection}{\Alph{section}}    
\setcounter{section}{0}

\section{Appendix: Technical details and proofs}
 \subsection{Construction of the ops graph} \label{sec:constr_ops_graph}
 In order to achieve polynomial complexity of the ops graph, we represent only operations in the graph which are $\mathcal{N_{BS-R}}$-valid for the given  neighborhood $\mathcal{N_{BS-R}}(x,p)$. Therefore, we introduced the concept of $\mathcal{N_{BS-R}}$-valid states in Section \ref{sec:ops_general} and limited the ops graph to $\mathcal{N_{BS-R}}$-valid states only. The goal of this section is to prove Lemma \ref{lem:valid_trans} from Section \ref{sec:ops_general} which provides instruction how to construct the ops graph in a forward induction. 
 
 The example from Figure \ref{fig:invalid_op} provides an intuition for identifying $\mathcal{N_{BS-R}}$-valid states. Lemma \ref{lem:interimset} formalizes this intuition by introducing a property which limits the possible sets of visited destinations inside of an operation which is relevant for $\mathcal{N_{BS-R}}(x,p)$.

 \begin{lemma}\label{lem:interimset}
 For the BS-R neighborhood $\mathcal{N_{BS-R}}(x,p)$, consider a $\mathcal{N_{BS-R}}$-valid operation $o=wsw'$ with $\{s\}=S$ and let define $m:=\min\{j: v_j\in S\}$ (the lowest index of destinations in $S$) and $M:=\max\{j: v_j\in S\}$ (the largest index of destinations in $S$).
Then
 \begin{align}
 \forall i \in [m+p,M-p] \text{ we have } v_i\in S \label{eq:ops2}
 \end{align}
 \end{lemma}

 \proof
 Consider a $\mathcal{N_{BS-R}}$-valid operation $o\in wsw'$ with $\{s\}=S$ and $m$ and $M$ as defined above. 
 
 We prove Condition~(\ref{eq:ops2}) by the law of contraposition. Let $v_l\in \left(V_d\setminus S\right)$. Then, for any drone tour $\pi_d\in \mathcal{N_{BS-R}}(x,p)$ with $o\in \pi_d$, two cases are possible. Either $v_l$ precedes the operation $o$ in $\pi_d$, then $l < m+p$ by the precedence constraints (\ref{eq:bs}) of the Balas-Simonetti neighborhood, or $v_l$ follows the operation $o$ in $\pi_d$, then $l> M-p$ by the precedence constraints (\ref{eq:bs}) of the Balas-Simonetti neighborhood. We conclude that $v_i\in S$ $\forall i \in [m+p,M-p]$. 

Proposition~\ref{prop:OP_BS} uses this lemma to give a full characterization $\mathcal{N_{BS-R}}$-valid states in the ops graph. Note that all initial states are $\mathcal{N_{BS-R}}$-valid by definition.

 \begin{proposition}\label{prop:OP_BS}
 For the BS-R neighborhood $\mathcal{N_{BS-R}}(x,p)$, consider state $(w,S,v)$ with $w\in V_r$, $S\subseteq V_d, |S|\geq 1$, and $v\in S\cup V_r$. Define indices $m$ and $M$ as in Lemma \ref{lem:interimset}.
 Then $(w,S,v)$ is $\mathcal{N_{BS-R}}$-valid iff one of the following is true:
 \begin{itemize}
 \item if $v\in V_r$ then 
 \begin{align}
 - \qquad \forall l \in [m+p, M-p ]: \ v_l \in S  \label{eq:term_valid}
\end{align}
\item if $v=v_i\in S$ then
 \begin{align}
 &- \qquad \forall l \in [m+p, M-p ]: \ v_l \in S  \label{eq:interim_valid1} \\
&- \qquad\forall v_l \in S: \ i>l-p  \label{eq:interim_valid2}
 \end{align}
 \end{itemize}
 \end{proposition}
 \proof
 The necessity of Conditions~(\ref{eq:term_valid}) and (\ref{eq:interim_valid1}) follows directly from Lemma~\ref{lem:interimset} and the necessity of Conditions~(\ref{eq:interim_valid2}) as a consequence of the Balas-Simonetti precedence constraints (\ref{eq:bs}).

 Let's consider the sufficiency of Conditions~(\ref{eq:interim_valid1}) and (\ref{eq:interim_valid2}) for the case of $v\in S$, since the proof can be directly extended to the remaining case. 

 Define set $T_1$ as the subset of $V_d\setminus S$ such that $\forall v_j \in T_1: j<m+p$ and $T_2$ as the subset of $V_d\setminus S$ such that $\forall v_j \in T_2: j>M-p$. Because of Condition~(\ref{eq:interim_valid1}), we have $T_1 \cup S \cup T_2=V_d$.  

 Let's define $\langle T\rangle$ as a sequence of destinations of some set $T\subseteq V_d$ ordered in increasing order of their indices. Consider the sequence of destinations $\rho=\left(\langle T_1\rangle,\langle S\setminus\{v\}\rangle,v,\langle T_2\rangle\right)$. Observe that:
 \begin{itemize}
 \item $\forall v_j\in T_1, v_l\in S$: we have $j<l+p$ and Balas-Simonetti precedence constraints (\ref{eq:bs}) hold.
 \item $\forall v_l\in S\setminus \{v=v_i\}$, $v_i$: we have by condition (\ref{eq:interim_valid2}) $i>l-p$ and Balas-Simonetti precedence constraints (\ref{eq:bs}) hold.
 \item $\forall v_j\in S, v_l\in T_2$: we have $l>j-p$ and Balas-Simonetti precedence constraints (\ref{eq:bs}).
 \end{itemize}

 Since constraints~\ref{eq:bs} applies straightforwardly to the remaining cases by construction, we conclude that $\rho\in\mathcal{N_{BS}}(x,p)$. We can construct a drone tour $\pi_d\in\mathcal{N_{BS-R}}(x,p)$ with operation $o=(w,\langle S\setminus\{v\}\rangle,v,w')\in\pi_d$ for some arbitrary $w'\in V_r$ and $\pi_d(V_d)=\rho$ accordingly. Thus state $(w,S,v)$ is $\mathcal{N_{BS-R}}$-valid.  
 \endproof
 Now we are able to proof Lemma \ref{lem:valid_trans}, we which repeat here:
 
\begin{lemma} \label{lem:valid_trans_app}
For the BS-R neighborhood $\mathcal{N_{BS-R}}(x,p)$ with $p\in \mathbb{N}$, consider non-terminal $\mathcal{N_{BS-R}}$-valid state $(w,S,v)\in \mathcal{O}$ with $|S|\geq 1$. Consider numbers $m:=\min\{j:v_j\in S\}$  and $M:=\max\{j:v_j\in S\}$. Then,
a corresponding state $(w,S\cup\{v'\},v'=v_i)$ with $v'\in V_d\setminus S$ is $\mathcal{N_{BS-R}}$-valid iff
\begin{align}
\begin{cases} i\in[M-p+1, \max\{M-1, m+2p-1\}] & \text{ or }\\ i\in[\max\{M+1, m+2p\},M+p] \text{ and } \left(v_j\in S \text{ } \forall j\in[m+p,i-p]\right) \end{cases} \label{eq:destseq}
\end{align}
Each corresponding terminal state $(w,S,w')$ with $w'\in V_r$ is $\mathcal{N_{BS-R}}$-valid.

\end{lemma}

\proof
The second statement follows straightforwardly. For the first statement, define $M':=\max\{M,i\}$ and $S'=S\cup\{v'\}$. The necessity of Condition~(\ref{eq:destseq}) for state $(w,S',v')\in \mathcal{O}$ follows from the following analysis of the values of $i$ in $v'=v_i$:
\begin{itemize}
\item $i\leq M-p$: impossible given Condition~(\ref{eq:interim_valid2}) of Proposition~\ref{prop:OP_BS}.
\item $i=M$: impossible since $v_i\in V_d\setminus S$.
\item $i\in [M+p+1,n_d]$: impossible since in this case there is $v_l  \in V_d \setminus S'$ with $l\in[M+1,M'-p]$, in contradiction with Condition~(\ref{eq:interim_valid1}) of Proposition~\ref{prop:OP_BS}.
\item $i\in [\max\{M+1, m+2p\},M+p]$: In this case, $M'=i$ and $[m+p,M'-p]\neq \emptyset$. Therefore, Proposition~\ref{prop:OP_BS} applies to $(w,S',v')$ only if $(v_j\in S \text{ } \forall j\in[m+p,i-p])$.
\end{itemize}

We proof the sufficiency of Condition~(\ref{eq:destseq}), by distinguishing the following cases:
\begin{itemize}
\item $i\in [M-p+1,M-1]$: Since $M'=M$ and $(w,S,v)\in \mathcal{O}$, Proposition~\ref{prop:OP_BS} applies to $(w,S',v')$.
\item $i\in [M+1,m+2p-1]$: Since $M'=i$ and $[m+p,M'-p]=\emptyset$, Proposition~\ref{prop:OP_BS} applies to $(w,S',v')$.
\item $i\in [\max\{M+1, m+2p\},M+p]$: Since $M'=i$, Proposition~\ref{prop:OP_BS} applies to $(w,S',v')$.  
\end{itemize}

\endproof
 \subsection{Complexity of the ops graph}\label{sec:app_ops_complexity}
The goal of this section is to prove the upper bound for the polynomial time complexity of the ops graph $\mathcal{O}$ from Proposition \ref{prop:comp_ops} in Section \ref{sec:ops_complexity}. We outlined that the complexity is linear to the amount of transition arcs in $\mathcal{O}$, and that the number of outgoing arcs of non-terminal states is bounded by $n_r+2p-1$, while terminal states have no outgoing arcs. What remains to complete the proof of Proposition \ref{prop:comp_ops} is to give approximate the total amount of non-terminal states in $\mathcal{O}$.

\begin{lemma}\label{lem:nbr_states_first}
 Let $U$ be the number of subsets $S\subseteq V_d$ with $\vert S \vert = k>1$, $n_d\geq 4p-2$ (large enough number of destinations), $m:=\min\{j:v_j \in S\}$ and $M:=\max \{j : v_j \in S\}$ such that Condition~(\ref{eq:interim_valid1}) from Proposition~\ref{prop:OP_BS} holds, i.e:
 \begin{align}
     \forall l \in [m+p,M-p]: v_l \in S \label{eq:interval}
 \end{align}
 Then:
 \begin{align}
     &\bullet \qquad \text{ if } k \in [2p,n_d-2p+2]\text{, then } U=(n_d-k-p+2)4^{p-1}\label{eq:lem4_if1}\\
    & \bullet \qquad \text{ if }k<2p \text{, then }U \leq (n_d-k-p+2)4^{p-1} \label{eq:lem4_if2} \\
     & \bullet \qquad \text{ if }k> n_d-2p+2\text{, then } U \leq p\cdot 4^{p-1} \label{eq:lem4_if3}
 \end{align}
 \end{lemma}

 \proof
Let construct a subset $S$ as described above. For any choice of the  minimal index $m$ and maximal index $M$, let denote by $T_{m,M}$ the subset of all destinations with indices between $m$ and $M$: $T_{m,M}=\{v_m, v_{m+1}, ...v_{M-1},v_{M} \}$, and let denote $l_{m,M}$ its length: $l_{m,M}=M-m+1.$  

In the following, we will count possible ways to select $T_{m,M}$ from $V_d$ and $S$ from $T_{m,M}$ for each possible $k=|S|$.

Since $S$ needs to accommodate $k$ destinations, and $S \subseteq T_{m,M}$, we must have: $l_{m,M}\geq k$. On the other hand, $T_{m,M}\subseteq V^d$, thus: $l_{m,M}\leq n_d$. 

\begin{itemize}
\item[i)] The desired subset $S$ is completely defined by the choice of $m$, $M$ and $T_{m,M} \setminus S$. According to Condition~(\ref{eq:interval}), $\forall v_j \in  T_{m,M} \setminus S$, we must have:
\begin{align*}
 j\in \ I_{m,M}: = ]m,\min\{m+p,M\}[ \ \cup \ ]\max\{M-p,m\},M[,   
\end{align*}
with $2p-2$ being the maximal length of $I_{m,M}$. 
\item[ii)] All together,  $m$ and $M$ must be chosen such that:
\begin{align*}
k \leq l_{m,M} \leq\min\{|S|+|I_{m,M}|, n_d\}\leq \min \{k+2p-2, n_d\}
\end{align*}

\item[iii)] Moreover, for fixed length $l_{m,M}$, there are exactly $n_d-l_{m,M}+1$ subsets $T_{m,M}$ of that length in $V_d$.
\end{itemize}

Case 1: $2p \leq k \leq n_d-2p+2$.

Since $l_{m,M}\geq k\geq 2p$, then $M-m+1\geq 2 p$ as well as the interval $I_{m,M}$ is exactly $[m+1,m+p-1]\cup[M-p+1,M-1]$ and has $2p-2$ elements. 

So, for given $m,M$ with associated length $l_{m,M}=k+i$, and $i \in [0,\min\{2p-2, n_d-k\}]=[0,2p-2]$, we have $\binom{2p-2}{i}$ possibilities to select the indices of $T_{m,M}\setminus S$ out of $I_{m,M}$.

Putting i), ii) and iii) together, the amount $U$ of choices for $S$ with $\vert S \vert =k$, is given by:
\begin{align}
    U=&\sum\limits_{i=0}^{2p-2} (n_d-(k+i)+1) \binom{2p-2}{i} \label{eq:upper_bound}\\
   =& (n_d-k+1) \sum\limits_{i=0}^{2p-2} \binom{2p-2}{i}- \sum\limits_{i=0}^{2p-2}i \binom{2p-2}{i} \nonumber \\
   =& (n_d-k+1)2^{2p-2}-(2p-2)2^{2p-2-1} \nonumber \\
   =& (n_d-k-p+2)4^{p-1} \nonumber
\end{align}
Case 2: $k < 2p$.

In this case, it is possible to select $m$ and $M$ such that $l_{m,M}<2p$, which implies that the left sub-interval and right sub-interval of $I_{m,M}$ overlap. The size of $I_{m,M}$ reduces then from $2p-2$ to $l_{m,M}-2$. 

Thus, we need to split up the number of choices $U$ in two sums, according to the distance of the selected $m$ and $M$. Note that  $l_{m,M}$ can still take all values $k+i$ for $i \in [0,2p-2]$.
\begin{align*}
    U=&\sum\limits_{i=0}^{2p-1-k} (n_d-(k+i)+1) \binom{k+i-2}{i} + \sum\limits_{i=2p-k}^{2p-2} (n_d-(k+i)+1) \binom{2p-2}{i}\\
\end{align*}    
The amount $U$ is obviously bounded by the expression on the right hand side from Equation~(\ref{eq:upper_bound}). 

Case 3: $k = n_d-2p+2+j$ for $j\in [1,2p-2]$.

Observe that $k+2p-2-j=n_d$. In this case, $m$ and $M$ must be chosen such that $l_{m,M}\leq n_d= k+2p-2-j$. The remaining construction of $U$ is identical to case 1.
\begin{align*}
U & =  \sum\limits_{i=0}^{2p-2-j}(n_d-(k+i)+1) \binom{2p-2}{i} \\ 
& \leq \sum\limits_{i=0}^{2p-2-j}(n_d-(k+i)+1+j) \binom{2p-2}{i} \\
& \leq \sum\limits_{i=0}^{2p-2} (n_d-(k-j+i)+1) \binom{2p-2}{i} \\
& =p\cdot 4^{p-1} \text{    by Formula (\ref{eq:upper_bound})} 
\end{align*}
 \endproof

Using the technical upper bound from Lemma~(\ref{lem:nbr_states_first}), we can now provide an approximation for the number of non-terminal states in $\mathcal{O}$:

\begin{lemma}\label{lem:nbr_states}
 Consider the BS-R neighborhood $\mathcal{N_{BS-R}}(x,p)$. The associated ops graph $\mathcal{O}$ has $O(n_d^2  n_r p \cdot 4^{p})$ non-terminal states.
 \end{lemma}

 \proof
By construction, there are $n_d\cdot n_r$ non-terminal states at stage 1, each having transitions to at most $2p-1$ different non-terminal states at stage 2 by Lemma~\ref{lem:valid_trans}. Thus there are $O(n_d\cdot n_r(2p-1))$ non-terminal states at stage 2. 
For any state $(w,S,v), v\in  S$ at stage $k=|S|\geq 3$:
\begin{itemize}
\item $w$ can take $n_r$ possible values
\item set $S\setminus \{v\}$, which has $k-1$ items, can take the following number of values (insert $\vert S\setminus \{v \} \vert =k-1$ in equations (\ref{eq:lem4_if1})-(\ref{eq:lem4_if3}) of Lemma~\ref{lem:nbr_states_first}): 
  \begin{itemize}
  \item not more than $(n_d-k-p+3)4^{p-1}$ values if $k\leq n_d-2p+3$
  \item not more than $p\cdot 4^{p-1}$ values otherwise 
  \end{itemize}
\item $v$, provided fixed $S\setminus \{v\}$, can take not more than $2p-1$ values according to Lemma~$\ref{lem:valid_trans}$
\end{itemize}
Observe that there are $(2 p-3)$ stages with $k>n_d-2p+3$. In total, for all stages $k\geq 3$, the number of non-terminal states is not larger than:
\begin{align*}
    &(2p-3)n_r(2p-1)p \cdot 4^{p-1} + \sum\limits_{k=3}^{n_d-2p+3} n_r (2p-1)  (n_d-k-p+3)4^{p-1} \\
    =\ &n_r(2p-1)\cdot 4^{p-1}\cdot [\ (2p-3)p + \sum\limits_{k=0}^{n_d-2p}(n_d-k-p)] \\
    = \ &n_r(2p-1)\cdot 4^{p-1}\cdot [\ (2p-3)p + (n_d-2p+1) \frac{1}{2}n_d  ] \\
    = \ &O\left(p \cdot n_r  \cdot 4^{p} \cdot[(n_d)^2 + p^2 - p\cdot n_d ]\right) \\
    = \ &O\left(p \cdot n_d^2  n_r  \cdot 4^{p}\right) 
\end{align*}

 \endproof
Multiplying the amount of non-terminal states in $\mathcal{O}$ with the upper bound of outgoing transition arcs for each of those states, provides us with the complexity of the ops graph as stated in Proposition \ref{prop:comp_ops}.

\subsection{Construction and complexity of the meta graph}
\label{sec:app_meta}
The meta graph implicitly enumerates all drone tours in a given BS-R neighborhood $\mathcal{N_{BS-R}}(x,p)$. In order to achieve polynomial complexity of the meta graph, we reduce it to so-called $\mathcal{N_{BS-R}}$-valid metastates and transitions which are relevant for $\mathcal{N_{BS-R}}(x,p)$, while preserving the general architecture of the graph described in Section \ref{sec:meta_general}. In this section, we prove Proposition \ref{prop:valid_encoding} and Proposition \ref{prop:metaarcs}, which characterize the special structure of valid metastates and transitions, respectively, given the new encoding scheme introduced in Section \ref{sec:meta_complexity}.
We repeat and prove Proposition \ref{prop:valid_encoding} below.
\begin{proposition} \label{prop:app_valid_encoding}
For the BS-R neighborhood $\mathcal{N_{BS-R}}(x,p)$, the encoded state $(S^-_k, S^+_k,w)$ is a $\mathcal{N_{BS-R}}$-valid metastate at stage $k\in[n_d]$ if and only if the following conditions hold:
\begin{align}
& - \quad |S^-_k|=|S^+_k| \leq \frac{p}{2} \label{eq:k_cardinality} \\
& - \quad S^-_k\subseteq\{k+1, \ldots,k+p-1\} \text{, }S^+_k\subseteq\{k-p+2, \ldots,k\}   \label{eq:k_candidates} \\
& - \quad \max\{l:l\in S^-_k\}-\min\{h:h\in S^+_k \} < p  \label{eq:k_distance}
\end{align}

\end{proposition}
\proof
The necessity of the left hand side of Condition~(\ref{eq:k_cardinality}) follows straightforwardly from the definition of  sets $S^-_k$ and $S^+_k$.  We refer to \cite{balas1999} for the proof of necessity for the remaining conditions.

Now suppose that for two sets $S^-_k$ and $S^+_k$ as defined by Equations~(\ref{eq:encoding}),  Conditions~(\ref{eq:k_cardinality})-(\ref{eq:k_distance}) are satisfied. We can define a corresponding permutation $\sigma$ of $[n_d]$, such that the BS precedence constraints for $x'=(x_{\sigma^{-1}(1)}, x_{\sigma^{-1}(2)}, ..., x_{\sigma^{-1}(n_d)})$ hold.

Set $ \sigma(i) = i $ for $i \in [n_d]\setminus (S^-_k \cup S^+_k)$ and consider any bijective mapping from $S^-_k$ to $S^+_k$.  Let check precedences for all possible configurations of a pair of indices $i,j \in [n_d]$ with $i+p\leq j$:
\begin{itemize}
    \item $i,j \notin S^-_k \cup S^+_k$: by definition $\sigma(i)=i<j=\sigma(j)$
    \item $i \notin S^-_k \cup S^+_k$ and $j \in S^-_k$: then by definition and Condition~(\ref{eq:k_distance}), \\ $\sigma(j) \geq \min \lbrace h: h \in S^+_k  \rbrace > \max \lbrace l : l \in S^-_k \rbrace -p \geq j-p \geq i = \sigma(i) $
    \item $i \notin S^-_k \cup S^+_k$ and $j \in S^+_k$: then $i \leq k$ and $\sigma(j)>k\geq i=\sigma(i)$ by definition
    \item $i\in S^+_k$: then we must have $j\notin S^-_k \cup S^+_k$ because of Conditions~(\ref{eq:k_candidates}) and~(\ref{eq:k_distance}). So,\\ $\sigma(i)\leq \max \lbrace l: l\in S^-_k \rbrace < \min \lbrace h: h \in S^+_k \rbrace +p  \leq  i+p \leq j = \sigma (j)$
    \item $i\in S^-_k$: then  $j >k$ and $j \notin S^-_k \cup S^+_k$ because of Conditions~(\ref{eq:k_candidates}) and (\ref{eq:k_distance}). So,\\ $\sigma(i)\leq k < j = \sigma(j)$
\end{itemize}
In Section~\ref{sec:formal} we assumed that the maximal flight time $e_{max}$ is high enough to  visit each destination in a return flight from its closest-by RL. 
Since state $(S^-_k, S^+_k, w)$ is relevant for this $\pi_d$, it is  $\mathcal{N_{BS-R}}$-valid. 
\endproof
Let's recall Proposition \ref{prop:metaarcs}.
 \begin{proposition} 
 There is an arc between metastate $(S^-_k,S^+_k,w) \in \mathcal{V}$ at stage $k$  and metastate $(S^-_{k+h},S^+_{k+h},w')\in \mathcal{V}$ at stage $k+h, h\geq0$  if and only if one of the cases below holds:
 \begin{itemize}
 \item $h=0$ and $S^-_{k+h}=S^-_k$, $S^+_{k+h}=S^+_k$
 \item $h\geq 1$, operation set $O=wHw'$ that corresponds to the transition arc belongs to $\mathcal{OP}^*_{BS-R}$ and the following conditions hold simultaneously:  
     \begin{align}
     & - \quad \lbrace j \in S_k^-: j> k+h \rbrace \subseteq S^-_{k+h}  \label{eq:metaarc_cond1} \\
     & - \quad \lbrace j\in S^-_k:j\leq k+h \rbrace \cap S^+_{k+h}= \emptyset  \label{eq:metaarc_cond2} \\
     & - \quad \lbrace j\in S^+_{k+h} : j \leq k \rbrace \subseteq S^+_k  \label{eq:metaarc_cond3}
     \end{align}
 \end{itemize}

Thereby $H=\{v_l \ : \ l\in I\}, \ I= (\lbrace k+1, ..., k+h \rbrace \cup S^+_k \cup S^-_{k+h}) \setminus (S^+_{k+h} \cup S^-_k)$.

 \end{proposition}
\proof
The case of $h=0$ is straightforward.

Consider $h\geq 1$. Let $S=[k]\setminus S^+_k \cup S^-_k$ and $T=[k+h]\setminus S^+_{k+h} \cup S^-_{k+h}$ be the index sets of the $k$ first visited and $k+h$ first visited destinations respectively. By the definition of transition arcs in Section~\ref{sec:meta_general}, the following should be true: $S\subset T$. 

Let show that $S\subset T$ is equivalent to Conditions~(\ref{eq:metaarc_cond1})-(\ref{eq:metaarc_cond3}).

Given $j\in S$, let show that $j\in T$ by analysing the following cases:
\begin{itemize}
    \item $j\leq k$: Then $j \notin S^+_k$ (by definition of $S$), $j\notin S^+_{k+h}$ because of Condition~(\ref{eq:metaarc_cond3}). Thus, $j\in T$.
    \item $k < j \leq k+h$: Since $j\in S$, we have $j \in S^-_k$. Then $j\in T$ because of Condition~(\ref{eq:metaarc_cond2}).
    \item $j > k+h$: Since $j\in S$, we have $j \in S^-_k$.  Then $j\in T$ because of Condition~(\ref{eq:metaarc_cond1}).
\end{itemize}
The necessity of conditions Conditions~(\ref{eq:metaarc_cond1})-(\ref{eq:metaarc_cond3}) if $S\subset T$ follows straightforwardly. 

Now, we need a formula to deduce the operation set $O=wHw'$ corresponding to the transition arc  between metastate $(S^-_k,S^+_k,w) \in \mathcal{V}$ at stage $k$ and metastate $(S^-_{k+h},S^+_{k+h},w')\in \mathcal{V}$ at stage $k+h, h\geq 1$, with the new encoding. 
Let $I_k=[k]\setminus S^+_k \cup S^-_k$ and $I_{k+h}=[k+h]\setminus S^+_{k+h} \cup S^-_{k+h}$ be the index sets of the $k$ first visited and $k+h$ first visited destinations, respectively.  

We can compute the set of destinations $H$ as  $H=\{v_l \ : \ l\in I\}$, with
\begin{align}
& 
I= I_{k+h} \setminus I_{k} = (\lbrace k+1, ..., k+h \rbrace \cup S^+_k \cup S^-_{k+h}) \setminus (S^+_{k+h} \cup S^-_k) \label{eq:seteq}
\end{align}
Note that we got the second equality of (\ref{eq:seteq}) by applying the well-known set properties
\begin{itemize}
\item[(1)] $A \setminus (B \cup C) = A \setminus B \setminus C = A \setminus C \setminus B$ 
\item[(2)]$ (A \cup B) \setminus C = (A \setminus C) \cup (B\setminus C) $ 
\item[(3)] $  A \setminus (B \setminus C) = (A \setminus B) \cup (A \cap C) $ 
\end{itemize}
to $I_{k+h}$ and $I_k$ from the left-hand side.
\endproof

Given the characteristics of $\mathcal{N_{BS-R}}(x,p)$-valid metastates from Proposition \ref{prop:valid_encoding} (or, equivalently, Proposition \ref{prop:app_valid_encoding}), Proposition \ref{prop:nbr_metastates} counts the amount of these states in the metagraph. The proof also sketches the construction for these states at fixed stage $k$ which we use to fill in the lookup table (c.f. Table \ref{tab:lookup}).
 \begin{proposition} \label{prop:nbr_metastates}
 For a given sequence of destinations $x$ and a given parameter $p$, meta  graph $\mathcal{G}$ for  neighbourhood $\mathcal{N_{BS-R}}(x,p)$  has $O(n_dn_r2^{p})$ states.
 \end{proposition}
 \proof
Consider state $(S^-_k, S^+_k,w)$ with $i:=|S^-_k|=|S^+_k|\geq \frac{p}{2}$ at some \textit{typical} stage $k$ (i.e., neither the first $(p-1)$ nor last $(p-1)$ stages). If $i=0$, there is exactly one possible pair of sets $S^-_k=S^+_k=\emptyset$. For $1 \leq i \leq \lfloor \frac{p}{2} \rfloor $, the rules of Proposition~\ref{prop:valid_encoding} allow us to construct all $\mathcal{N_{BS-R}}$-valid meta states in the following manner:

Let denote $m:=\max\{l:l\in S^-_k\}$. $S^-_k$ and $S^+_k$ can be chosen as any $i$-sized subsets of $\lbrace k+1, k+2, ..., m \rbrace$ that include $m$, and $\lbrace m-p+1,..., k-1, k \rbrace$, respectively.\\
To ensure that $S^-_k, S^+_k$ fit $i$ elements, the range of possible values for $m$ is given by: $k+i \leq m \leq k+p-i$.\\
Putting all together, the amount of $\mathcal{N_{BS-R}}$-valid metastates at stage $k$ equals exactly:

\begin{align}
    n_r\left(1+\sum\limits_{i=1}^{\lfloor \frac{p}{2} \rfloor} \sum\limits_{m=i}^{p-i} \binom{m-1}{i-1} \binom{p-m}{i}\right)  = n_r2^{p-1} \label{eq:bin_formula},
\end{align}

where Equality~(\ref{eq:bin_formula}) can be proven in a similar fashion as Lemma 4.6 in \citep{balas1999}.

For \textit{non-typical stages}, we can adapt a similar construction procedure of states, and further limit $S^-_k$ and $S^+_k$ to subsets of $[n_d]$. The statement of Proposition~\ref{prop:nbr_metastates} follows straightforwardly. 
 \endproof

\section{Appendix: Details on the case study}
\subsection{Drone equations}
\label{sec:AppDrone}
\begin{figure}
\centering
\includegraphics[width = 0.25\columnwidth]{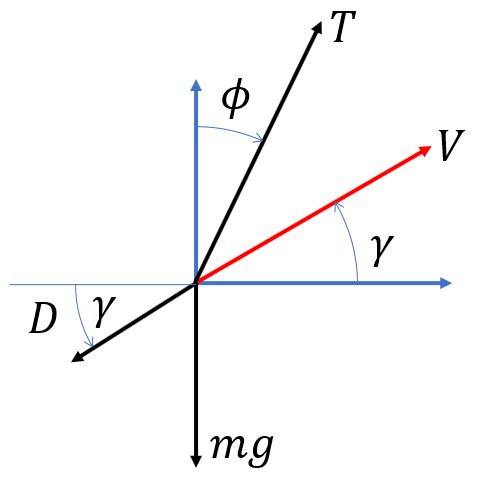}
\caption{Motion of a quadcopter \footnotesize{The motion of a quadcopter can be described as equilibrium of three forces: the thrust $T$, the weight $mg$ and the drag $D$. The drone tilts of $\phi$ to orient $T$ to counteract $mg$ and $D$.}} \label{fig:FlightBill} 
\end{figure}

Drones are well-known systems whose dynamics has been investigated for years, see \cite{quan2017introduction}. In this work the drone is conceived as a point of mass $m > 0$ subject to the gravity acceleration $g > 0$, the thrust force $T > 0$ and the drag force $D\in \mathbb{R}$ (which is always opposite the motion direction), see Figure \ref{fig:FlightBill}. The dynamics of the drone consists of the rotation of the body, namely with angle $\phi \in \mathbb{R}$, to orient the thrust in the desired motion direction. Assuming the drone to fly 
at constant speed $V > 0$, with a constant glide angle $\gamma \in \mathbb{R}$, the following equilibrium equation are formulated
\begin{subequations}
\label{eq:DroneEq}
    \begin{align}
    T \cos\phi &= mg + D\sin\gamma& \text{vertical equilibrium}\\
    T \sin\phi &= D\cos\gamma &\text{horizontal equilibrium}
    \end{align}
\end{subequations}
The use of standard aerodynamics arguments, see \cite{chattot2015theoretical}, let us describe the drag as $D= 1/2 \rho S V^2 C_D$ in which $\rho$ denotes the air density, $S$ is the drone cross section, and $C_D$ represents the drag coefficient.
From \eqref{eq:DroneEq} the computation of the thrust leads to
\begin{equation}
\label{eq:PropT}
    T = \sqrt{(mg)^2+2(mg) D\sin\gamma+D^2}.
\end{equation}
The propeller is conceived as an aerodynamic system which is fed by an input shaft torque $Q > 0$. The propeller performances are described by standard aerodynamics arguments, see \cite{chattot2015theoretical}, thanks to which we have
\begin{subequations}
\label{eq:PropAero}
\begin{align}
    T &= c_T \rho (w r)^2 (\pi r)^2\\
    Q &= c_Q \rho (w r)^2 (\pi r)^2 r
\end{align}
\end{subequations}
in which $c_T$ and $c_Q$ are dimensionless thrust and torque coefficients that depend on the geometry of the propeller,  $w >0$ is the propeller rotational speed, and $r$ represents the propeller radius. Moreover, the mechanic power required to rotate the propeller at $w$ is then computed as
\begin{equation}
\label{eq:PropPower}
\Pi = Q\,w.    
\end{equation}
On the other hand, the mechanic power is delivered to the shaft through an electric motor supplied by a battery with nominal voltage $v_b > 0$. Assuming the motor efficiency be $\eta \in (0, 1)$, the input-output power equilibrium leads to
\begin{equation}
\label{eq:MotorPower}
    v_b i_b \eta = \Pi
\end{equation}
in which $i_b > 0$ denotes the current delivered from the battery to the motor. Then, the current needed to fly at constant speed $V > 0$ is computed through the combination of  \eqref{eq:PropT}-\eqref{eq:MotorPower} as
\begin{equation}
    \begin{aligned}
 i_b &= \dfrac{1}{v_b\eta}\Pi\\
     &= \dfrac{1}{v_b\eta}Q\,w\\
     &= \dfrac{1}{v_b\eta} c_Q \rho (w r)^2 (\pi r)^2 r\,w\\
     &= \dfrac{1}{v_b\eta} \frac{c_Q  }{\pi r \sqrt{c_T^{3} \rho} } T^{3/2}\\
     &= \dfrac{1}{v_b\eta} \frac{c_Q  }{\pi r \sqrt{c_T^{3} \rho} } \left((mg)^2+2mgD\sin\gamma+D^2\right)^{3/4}\\
     &= \dfrac{1}{v_b\eta} \frac{c_Q  }{\pi r \sqrt{c_T^{3} \rho} } \left((mg)^2+2(mg)\left(\dfrac{1}{2}\rho S  C_D\right)\sin\gamma V^2+\left(\dfrac{1}{2} \rho S  C_D\right)^2 V^4\right)^{3/4}.
    \end{aligned}
\end{equation}
Finally, the notation is reduced through
\begin{subequations}
\begin{align}
    k &:= \dfrac{1}{v_b\eta} \frac{c_Q  }{\pi r \sqrt{c_T^{3} \rho} }\\
    \overline{C}_D &:= \left(\dfrac{1}{2} \rho S  C_D\right) \\
    W &:= mg
\end{align}
\end{subequations}
to obtain
\begin{equation}
\label{eq:current}
    \begin{aligned}
 i_b = k \left(W^2+2W\overline{C}_D\sin\gamma V^2+\overline{C}_D^2 V^4\right)^{3/4}.
    \end{aligned}
\end{equation}
It is worth to note that \eqref{eq:current} is valid for any flight condition at constant speed and glide angle. So, to compute the current $i_b$ needed to 
\begin{itemize}
    \item take off we impose $\gamma = \pi/2$
    \item flight at constant altitude we impose  $\gamma = 0$
    \item land we impose $\gamma = -\pi/2$.
\end{itemize}

\subsection{Generation of drone routes for search and rescue in practice} \label{sec:app_pract}

In order to conduct a comprehensive search of the area of interest in the woods around Lake Occhito (see Figure \ref{fig:fafb}), the first step is to define the set of aerial destinations through standard procedures exploited in vertical aerial photography \citep[see][]{Lear.1997}. This step provides the green cycles in Figure \ref{fig:fafb}. Secondly, adequate starting- and ending RLs $w_0$ and $w_t$ are defined, depending on the current position of the operator. In this scenario, those are RLs 12 and 100, respectively.

Both heuristics, VLSN-LS and PRACT require a visiting sequence of destinations as an input, which was selected as the shortest aerial TSP-path that connects $w_0$ and $w_t$ by visiting all destinations, generated with the Gurobi optimizer for Python in this case. 

Differently from VLSN-LS, the algorithm PRACT developed by practitioners proposes a solution that conserves the initial visiting sequence of destinations. Algorithm \ref{alg:HandMade} shows how PRACT plans the drone tour $\pi_d^{PRACT}$ in a greedy manner, by inserting replenishments to the visiting sequence of destinations. In the final step, the rover path is generated as the shortest ground path from $w_0$ to $w_t$ over all the replenishment nodes in $\pi_d^{PRACT}$.   

\begin{algorithm}
\caption{PRACT} \label{alg:HandMade}
\begin{algorithmic}
\State $E \leftarrow \xi_{max} $;\\ 
\State $\pi_d^{PRACT} \leftarrow$ Add $w_0$;\\ 
\Repeat
    \State $F \leftarrow$ Compute the flight bill to the next destination;  
    \State $R \leftarrow$ Compute the drone range (with landing) with $0.90\%E-F$; 
    \State $D \leftarrow$ Compute the distance from the next destination to the closest RL;\\ 
    \If{$D \leq R$}\\
      \State $\pi_d^{PRACT} \leftarrow$ Add the next destination;
      \State $ E \leftarrow E-F$;
      \Else
      \State $\pi^d \leftarrow$ Add the RL closest to the current destination.
      \EndIf \\
\Until{All destinations have been visited}\\
\Return $\pi_d^{PRACT}$      
\end{algorithmic}
\end{algorithm}
\newpage


\newpage

\end{document}